\documentclass[12pt,a4paper]{article}
\usepackage[utf8]{inputenc}
\usepackage{amsmath}
\usepackage{amsfonts}
\usepackage{amssymb}
\usepackage{graphicx}
\usepackage{color}
\usepackage{verbatim}
\usepackage{subfig}
\usepackage{authblk}
\usepackage{hyperref}
\usepackage{breqn}
\usepackage{geometry}
\providecommand{\keywords}

\makeatletter
\let\cat@comma@active\@empty
\makeatother
\begin{document}

\title{Quantum Berezinskii-Kosterltz-Thouless Transition for Topological Insulator}

\author[1,2]{Ranjith Kumar R}
\author[1,2]{Rahul S}
\author[3]{Surya Narayan}
\author[1]{Sujit Sarkar}
\affil[1]{Poornaprajna Institute of Scientific Research, 4, 16th Cross,
	Sadashivnagar, Bengaluru - 5600-80, India. }
\affil[2]{Manipal Academy of Higher Education, Madhava Nagar, Manipal - 576104, India.}
\affil[3]{Raman Research Institute, C. V. Raman Avenue, 5th Cross,
	Sadashivanagar, Bengaluru - 5600-80, India.}
\maketitle

\begin{abstract}
\noindent We consider the interacting helical liquid system at the one-dimensional edge of a two-dimensional topological insulator, 
coupled to an 
external magnetic field and s-wave superconductor and map 
it to an XYZ spin chain system. 
This model undergoes quantum 
Berezinskii-Kosterlitz-Thouless (BKT) transition
with two limiting conditions. 
We derive the renormalization group (RG) 
equations explicitly and also present the flow lines behavior. 
We also present the behavior of RG flow lines based on 
the exact solution. We observe that the physics of Majorana fermion 
zero modes and the 
gaped Ising-ferromagnetic phase, which appears in a different context. 
We observe that the evidence of gapless 
helical Luttinger liquid phase as a common 
non-topological quantum phase for both quantum BKT transitions. We explain analytically and physically that there is no Majorana-Ising transition.
In the presence of chemical potential, 
the system shows the commensurate to incommensurate transition.
\end{abstract}

\newpage
\noindent\textbf{Introduction}\\
Berezinskii \cite{berezinskii1971destruction}, 
in the year 1971 and Kosterlitz and Thouless 
\cite{KT-1973}, in the year 1973 have explained
a new kind of phase transition in two-dimensional 
XY spin model \cite{XY-model} using renormalization 
group (RG) method  
\cite{giamarchi2003quantum,2013ybkt.book...93O,senechal2004introduction,altland2010condensed,fradkin2013field,marino2017quantum}.
According to Mermin-Wagner-Hohenberg theorem 
\cite{mermin1966absence,PhysRev.158.383}, continuous symmetry cannot be broken spontaneously at any finite temperature for $d \le 2$ (d = dimension). This is because of strong fluctuations of the goldstone modes in $d=1,2$ which restore the broken symmetry at long distances for finite temparature.
However the classical XY model with $d=2$ is found to have power-law decay in correlation fucntion at low temparature and exponential decay in correlation fucntion at high temparature \cite{Nagaosa-QFT}. This predicted a new kind of phase transition between them, presently know as Berezinskii-Kosterlitz-Thouless (BKT) transition. \\
BKT transition can successfully explain the phase transition 
 in two-dimensional XY model by considering the topological 
 non-trivial vortex (topological defect) configuration, where 
 there is no requirement of spontaneous symmetry breaking. They have proposed that the disordering is facilitated 
 by the condensation of topological defects \cite{berezinskii1971destruction,KT-1973}. 
The basic explanation is that, at high temperature the correlation 
function decays exponentially and thermal generation of vortices is favorable 
for $T\geq T_c$ (where $T_c$ is critical temperature of BKT transition). Thus at higher 
temperature even number of vortices with 
opposite sign (i.e, vortex and anti-vortex) are produced 
and they are unbounded. At low temperature i.e, at $T<T_c$ 
the correlation function decays as power low and vortex 
and anti-vortex are bounded by forming a pair. Thus the phase transition takes place at 
the critical temperature which is obtained by minimizing the free energy \cite{jose2017duality}. 
This transition was first explained in two-dimensional XY model. 
Therefore the study of BKT transition is crucial in quantum many body systems 
since many quantum mechanical two-dimensional systems can be approximated 
to two-dimensional XY model \cite{timm2012theory}.\\
The physics of low dimensional quantum many body 
condensed matter system is enriched with its new and 
interesting emergent behavior. One-dimensional 
electronic systems cannot be solvable by the 
Fermi liquid theory due to the infrared divergence 
of certain vertices. An alternative theory called 
Tomonaga-Luttinger liquid theory has been 
constructed to describe the one-dimensional electronic 
system \cite{Haldane1981}. Hence we mention very 
briefly the nature of different Luttinger liquid physics 
to emphasis the enrich physics of helical Luttinger 
liquid. In this theory the Luttinger parameter ($K$) 
determines the nature of interaction. $K<1$ and $K>1$ 
characterizes the repulsive and attractive interactions 
respectively, where as $K=1$ characterizes non-interacting case \cite{helical1}. \\
The physics of Luttinger liquids (LL) can be of three 
different forms : spinful LL, chiral LL, and helical LL.
 Spinful LL shows linear dispersion around the Fermi level 
 with the difference of $2k_F$, in the momentum between 
 left and right moving branches. Chiral LL has spin 
 degenerated, strongly correlated electrons moving 
 in only one direction. In helical LL one can observe 
the Dirac point due to the crossing of left and right moving 
branches, also electrons with opposite spins move in opposite 
directions \cite{helical1}. In one dimension these helical 
edge states are protected by time-reversal (TR) symmetry 
with $T^2=-1$. In contrast to this, spinless LL satisfies 
$T^2=1$ and chiral LL breaks TR invariance in one 
dimension. Spinful LL has to have an even number of 
TR pairs where as helical LL can have an odd number 
of components \cite{wu2006helical}. \\The realization 
of spinful Luttinger liquids have been observed in 
carbon nanotubes \cite{LL-real2,LL-real3,LL-real4}, 
GaAs/AlGaAs heterostructures \cite{LL-real1}, 
cleaved edge overgrowth one-dimensional channel 
\cite{LL-real5}, which break the TR invariance, 
also chiral Luttinger liquids have been observed in 
fractional quantum Hall edge states \cite{LL-real6}. 
Quantum spin Hall insulator or topological insulator 
support the helical edge states, which are realized in HgTe 
\cite{LL-real7} and InAs/GaSb quantum wells \cite{helical1,LL-real8,LL-real9,LL-real10}. \\
Here we consider an interacting helical liquid system 
at the edge of the quantum spin Hall system as our 
model Hamiltonian. Quantum spin Hall systems with 
or without Landau levels describes the helical edge 
states and it also describes the connection between 
spin and momentum. 
The left movers in the edge of quantum spin Hall systems 
are associated with down spin and right movers with up spin 
\cite{PhysRevLett.95.146802,topo.insu,moore2010birth,nishimori2010elements,
PhysRevB.80.155131,PhysRevB.85.075125,PhysRevB.83.035107}.
In the non-interacting case the helical liquid is characterized 
by the $Z_2$ symmetry indicating that the even and odd TR 
components are topologically distinct \cite{PhysRevLett.95.146802}. 
In the interacting case it is observed that helical liquid with odd 
number of components can not be constructed in the one 
dimensional lattice \cite{wu2006helical}. Low-temperature 
conductance of a weakly interacting one-dimensional helical 
liquid without axial spin symmetry has been explored \cite{PELflex}. 
The formation of these one-dimensional states which can be 
controlled by the gate voltage on the topological surface has 
been studied and found the energy dispersion is almost linear 
in the momentum \cite{Nagosa}. The impact of interaction 
on the helical liquid system has been studied explicitly, 
which results in the forming of Mojorana fermion states 
with high degree of stability \cite{Altland-2011}. 
The scattering process between fermion bands conserving 
momentum of helical liquid system opens a gap against 
interaction effect, which leads to the stabilization of 
Majorana fermion mode \cite{Xu-Moore}. The existence 
of the Majorana fermion mode and the characterization of 
Majorana-Ising transition has also been studied extensively 
\cite{Sarkar-2016, Sudip-kumar}.\\ 
However the renormalization group study and the physics 
of quantum Berezinskii-Kosterlitz-Thouless (BKT) transition  
has not been studied explicitly for interacting helical edge 
states. Quantum BKT transition is a topological quantum 
phase transition in low dimensional quantum many-body 
system. But it has not been  explored explicitly in the 
literature for the helical edge states or for any quantum 
matter \cite{2013ybkt.book...93O,jose2017duality}. 
Therefore in this paper investigate the quantum BKT 
transition for the edge states of topological insulator. 
Quantum BKT transition happens at temparature $T=0$. 
Here we study how RG flow lines of the sine-Gordan 
coupling constant (i.e $B$ and $\Delta$ in the present problem)
behave with the LL parameter ($K$).

\noindent\textbf{Motivation and relevance of this study :} \\ 
{\bf First objective}: 
The physics of topological state of matter is the second
revolution of quantum mechanics \cite{haldane}. This important concept and new important
results with high impact not only bound to the general audience of different
branches of physics but also creates interest for the other branches of science.
This new revolution in quantum mechanics was honored by the Nobel
prize in physics in the year 2016.
This is
one of the fundamental motivation to study the topological state of matter 
for the edge state of topological insulator.\\
{\bf Second objective}:
One of main motivation of this study is to find the quantum 
BKT for the one-dimensional helical edge mode of a two-dimensional topological insulator and also the  limit in which it appears. 
We also search that what are the quantum 
phases appears in these quantum  BKT transition, and 
there is any relation between the quantum phases which 
appear in the two different quantum BKT transition.
There is no evidence of any Majorana-Ising transition for this quantum BKT transitions.\\ 
{\bf Third objective}:
The mathematical structure and results of the renormalization group (RG) theory are the
most significant conceptual advancement in quantum field theory in the last several decades
in both high-energy and condensed matter physics \cite{Zee-QFT}.
The need for the RG is more
transparent in condensed matter physics. Therefore in the present study, 
we use RG method to study the different quantum phases either topological or non-topological in character, through two quantum BKT Hamiltonians.\\ 
{\bf Fourth objective}:
It is very rare to find the exact solution for the problems of quantum 
many body condensed matter system. Here we find the exact solution 
of quantum BKT equation for the edge state of topological insulator. The 
other motivation is to find the exact solution for the RG 
flow line of this two quantum BKT equation.
Therefore the present study of quantum BKT provides 
a new perspective on topological quantum phase transition.\\
\textbf{\large Model Hamiltonian and the derivation 
of quantum BKT equations}\\
We consider the interacting helical liquid system at 
the one-dimensional edge of a topological insulator as our model system 
\cite{PhysRevLett.95.146802,bernevig-topo-insu,asboth2016short,
qi2011topological,wu2006helical}. 
These edge states are protected by the symmetries 
\cite{PhysRevB.85.075125,Classifi-symmetry}. Topological insulator is two-dimensional system but the physics of helical liquid at the edge of topological insulator is one-dimensional. In this edge states 
of helical liquid, spin and momentum are connected as the right 
movers are associated with the spin up and left movers 
are with spin down and vice versa. One can write the total fermionic field of the system as,
 \begin{equation}
 \psi(x)=e^{ik_F x} \psi_{R\uparrow}+e^{-ik_F x} \psi_{L\downarrow},
 \end{equation}
where $\psi_{R\uparrow}$ and $\psi_{L\downarrow}$ are 
the field operators corresponding to right moving (spin up) 
and left moving (spin down) electron at the both upper and 
lower edges of the topological insulators.\\
Here we discuss the basics of this model Hamiltonian 
very briefly \cite{wu2006helical,Altland-2011}.
For the low energy collective excitation in one-dimensional 
system one can write the Hamiltonian as,
\begin{equation} 
\begin{split}
H_0 &= \int \frac{dk}{2 \pi} {v_F} [ ({{\psi}_{R \uparrow}}^{\dagger} (i \partial_x)
{\psi_{R \uparrow} } -  {{\psi}_{L \downarrow}}^{\dagger} (i \partial_x)
{\psi_{L \downarrow} }) 
 +  ({{\psi}_{R \downarrow}}^{\dagger} (i \partial_x)
{\psi_{R \downarrow} } -  {{\psi}_{L \uparrow}}^{\dagger} (i \partial_x)
{\psi_{L \uparrow} })], 
\end{split}
\end{equation}
where the terms in the parenthesis represents 
Kramer's pair at both edges of the system.
The Hamiltonian for the non-interacting part of the 
one edge of the helical liquid system is,
\begin{equation}
 H_{01}={{\psi}_{L \downarrow}}^{\dagger}
 ({v_F}i \partial_x -\mu){\psi_{L \downarrow} }
 +{{\psi}_{R \uparrow}}^{\dagger}({-v_F}i \partial_x -\mu)
 {\psi_{R \uparrow}}.
\end{equation}
We consider the topological insulator in the proximity of 
s-wave superconductor ($\Delta$) and the magnetic field (B). 
Thus the additional part of the Hamiltonian is given by,
\begin{equation} 
\delta H =  \Delta {\psi_{L \downarrow}} {\psi_{R \uparrow}} + 
B {{\psi}_{L \downarrow}}^{\dagger}{\psi_{R \uparrow}} 
+ h.c . \end{equation} 
We will see in the present study that coupling $\Delta $ induce the topological superconducting phase and coupling $B$ induce the Ising-ferromagnetic phase.
One can find two types of interactions which are allowed by 
time-reversal in helical liquid system. They are Forward and 
Umklapp interactions \cite{laurell2012},
\begin{equation}H_{fw} = g_2  {\psi_{L \downarrow}}^{\dagger} 
{\psi_{L \downarrow}}
 {\psi_{R \uparrow}}^{\dagger} {\psi_{R \uparrow}} .\end{equation} 
\begin{equation} H_{um} = g_u {{\psi}_{L \downarrow}}^{\dagger} 
\partial_x {{\psi}_{L \downarrow}}^{\dagger}
{{\psi}_{R \uparrow}} \partial_x {{\psi}_{R \uparrow}} 
+ h.c.\end{equation} 
Thus we get the total Hamiltonian as, 
$ H=H_{01}+H_{fw}+H_{um}+\delta H$. The authors of ref.\cite{Altland-2011} have mapped this Hamiltonian to 
the XYZ spin-chain model (up to a constant) i.e, $H_{XYZ}= \sum_i H_i$, where
\begin{equation} H_i = \sum_{\alpha} J_{\alpha} {S_i}^{\alpha} {S_{i+1}}^{\alpha}
- [ \mu + B (-1)^i ] {S_i}^z  .\label{XYZ-spinform}\end{equation} 
This is our model Hamiltonian where, 
$J_x=v_F+\Delta$, $J_y=v_F-\Delta$ and $J_z=g_u$ 
are coupling constants.
One can write the model Hamiltonian in spinless fermion form after Jordan-Wigner transformation as \cite{Sudip-kumar},
\begin{multline}
H=\frac{-J}{2} \sum_i (c_i^{\dagger}c_{i+1}+h.c) + J_z \sum_i \left( c_i^{\dagger}c_{i}-\frac{1}{2}\right)  \left( c_{i+1}^{\dagger}c_{i+1}-\frac{1}{2}\right)\\ +\frac{\Delta}{2} \sum_i (c_{i+1}^{\dagger}c_{i}^{\dagger}+h.c) -\sum_i [\mu+B(-1)^i] \left( c_i^{\dagger}c_{i}-\frac{1}{2}\right) 
\end{multline}
After the continuum field theory, 
one can write the Hamiltonian as \cite{Altland-2011,Sarkar-2016,Sudip-kumar,SineG1,SineG2}, 
\begin{equation}
\begin{split}
H  = \frac{v}{2}  \int   \left[ \frac{1}{K}  {({\partial_x \phi(x) })}^2 + 
K {({\partial_x  \theta(x) })}^2  \right]  dx
 +  \frac{B}{\pi} \int  cos(\sqrt{4 \pi} \phi(x)) dx -  \frac{\Delta}{\pi} \int  cos(\sqrt{4 \pi }\theta(x)) dx\\
+ \frac{g_u}{2 {\pi}^2 } \int  cos(4 \sqrt{\pi} \phi(x)) dx-  \frac{\mu}{\sqrt{\pi}} \int  \partial_x \phi(x) dx, \label{model1}
\end{split}
\end{equation}
where $\theta(x)$ and $\phi(x)$ are the dual fields and $K$ 
is the Luttinger liquid parameter \cite{Luttinger-1963} 
of the system. 
\noindent The author of ref.\cite{Sarkar-2016}
has shown explicitly the $g_u$ has no effect on the
topological state and also on the Ising-ferromagnetic state of the system. 
Therefore the Hamiltonian is reduced to,
 \begin{equation}
 	\begin{split}
 	H  =  \frac{v}{2} \int  \left[ \frac{1}{K} ( {({\partial_x \phi(x) })}^2 + 
 	K {({\partial_x  \theta(x) })}^2  )\right] dx
 	+ \frac{B}{\pi} \int  cos(\sqrt{4 \pi} \phi(x)) dx - \frac{\Delta}{\pi}  \int  cos(\sqrt{4 \pi \theta(x)}) dx\\- \frac{\mu}{\sqrt{\pi}}\int   \partial_x \phi(x) dx.\label{model2}
 	\end{split}
 	\end{equation}
 	
\textbf{\large Quantum BKT equations}\\
The model Hamiltonians (eq.\ref{XYZ-spinform} and eq.\ref{model1}) have already been studied in different context 
in quantum spin systems and also in condensed matter field theory 
\cite{fradkin2013field,Sarkar-2016,Sudip-kumar}. 
In the present study, we are only interested in  
the physics of quantum BKT, which has not been studied in the literature.  We use quantum field theoretical renormalization group method to study this problem which predict and explain the enriched physics of quantum BKT in elegant way. 
The BKT equations can be derived by considering two 
limiting situations, i.e, one BKT equation for $B=0$ and other is for $\Delta=0$, in the Hamiltonian $H$ (eq.\ref{model2}).
This gives two model Hamiltonian $H_1$ and $H_2$ as follows,
\begin{equation} H_1 = \frac{v}{2}\int \left[ \frac{1}{K} 
(\partial_x \phi(x))^2 + K(\partial_x \theta(x))^2\right]  dx
- \frac{\Delta}{\pi}\int   \cos(\sqrt{4\pi}\theta(x)) dx  .\end{equation}
\begin{equation}
H_2 =\frac{v}{2}\int \left[ \frac{1}{K}(\partial_x \phi(x))^2 
+ K(\partial_x \theta(x))^2\right] dx+ \frac{B}{\pi} \int \cos(\sqrt{4\pi}\phi(x)) dx.
\end{equation} 
At first we set $\mu=0$ for simplicity and consider its effect in later section.\\
\textbf{ Results for Hamiltonian $H_1$}:\\
Here we present the quantum BKT equations for the 
Hamiltonian $H_1$ and show that at $T=0$ there exist 
different quantum phases with topological and 
non-topological properties and also the crossover 
between them,
\begin{equation} H_1 = \frac{v}{2}\int \left[ \frac{1}{K} 
(\partial_x \phi(x))^2 + K(\partial_x \theta(x))^2\right] dx  
- \frac{\Delta}{\pi}\int \cos(\sqrt{4\pi}\theta(x)) dx .\end{equation} 
Finally BKT equation can be derived for the Hamiltonian 
$H_1$ as (please see appendix A for detailed derivation),
\begin{equation}
\dfrac{d{\Delta}}{dl}= \left( 2-\frac{1}{K}\right) {\Delta}, \;\;\;\;\;\;\;
\frac{d K}{dl} =  {\Delta}^2 . \label{RGdelta}
\end{equation}         
To reduce eq.\ref{RGdelta} to standard form of BKT, 
we do the following transformations,
$-y_{||} =\left( 1 -  \frac{1}{2K}\right)  $ 
and finally the RG equations become,
\begin{equation} 
\frac{d \Delta}{dl} = - y_{||} \Delta ,\;\;\;\;\;\;
\frac{d y_{||}}{dl} = - {\Delta^2}. \label{RGdelta1}
\end{equation}
	\begin{figure}[h]
						\subfloat[]{\includegraphics[scale=0.43]{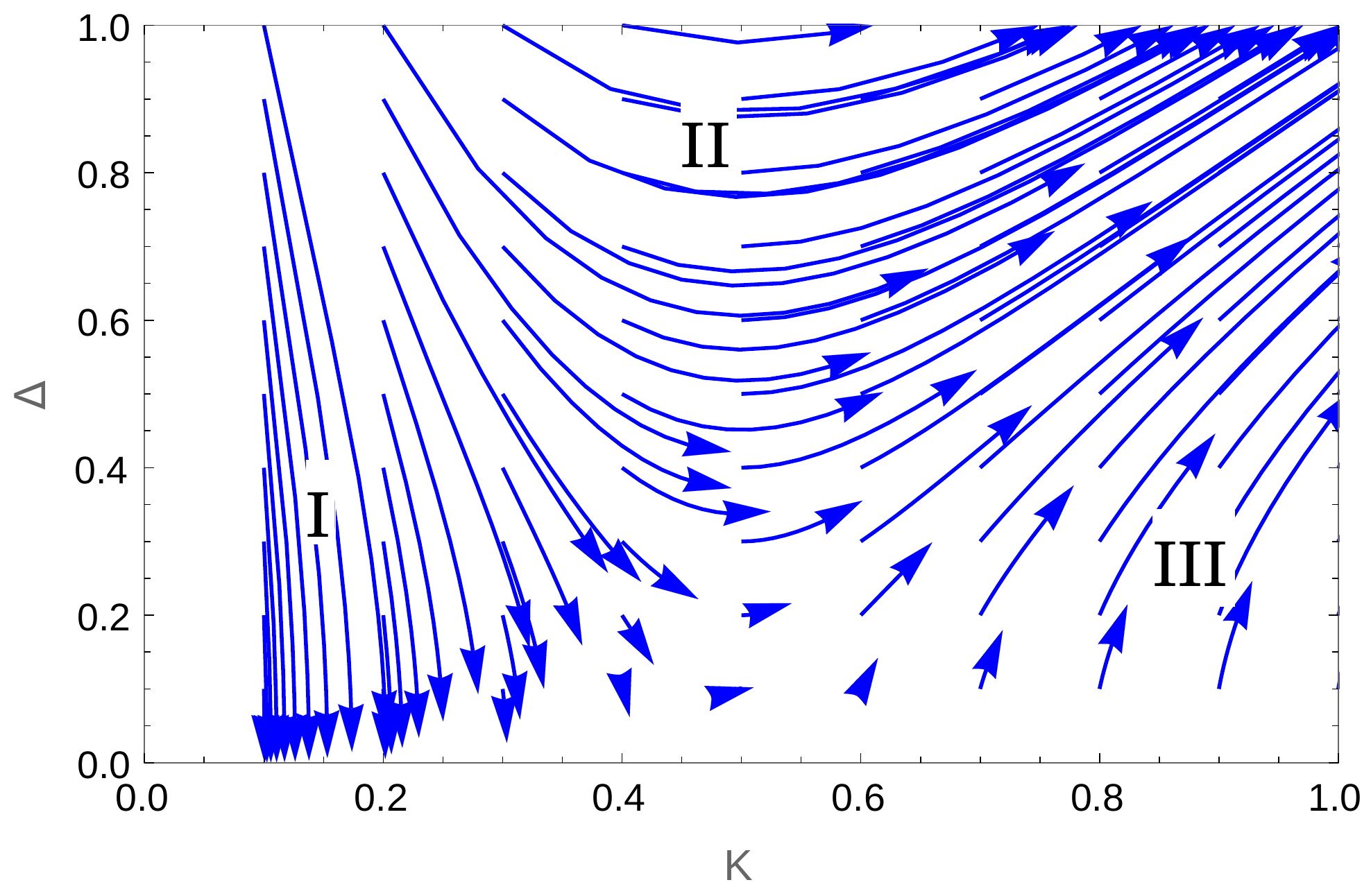}\label{RG1}}
						\;\;\;\subfloat[]{\includegraphics[scale=0.43]{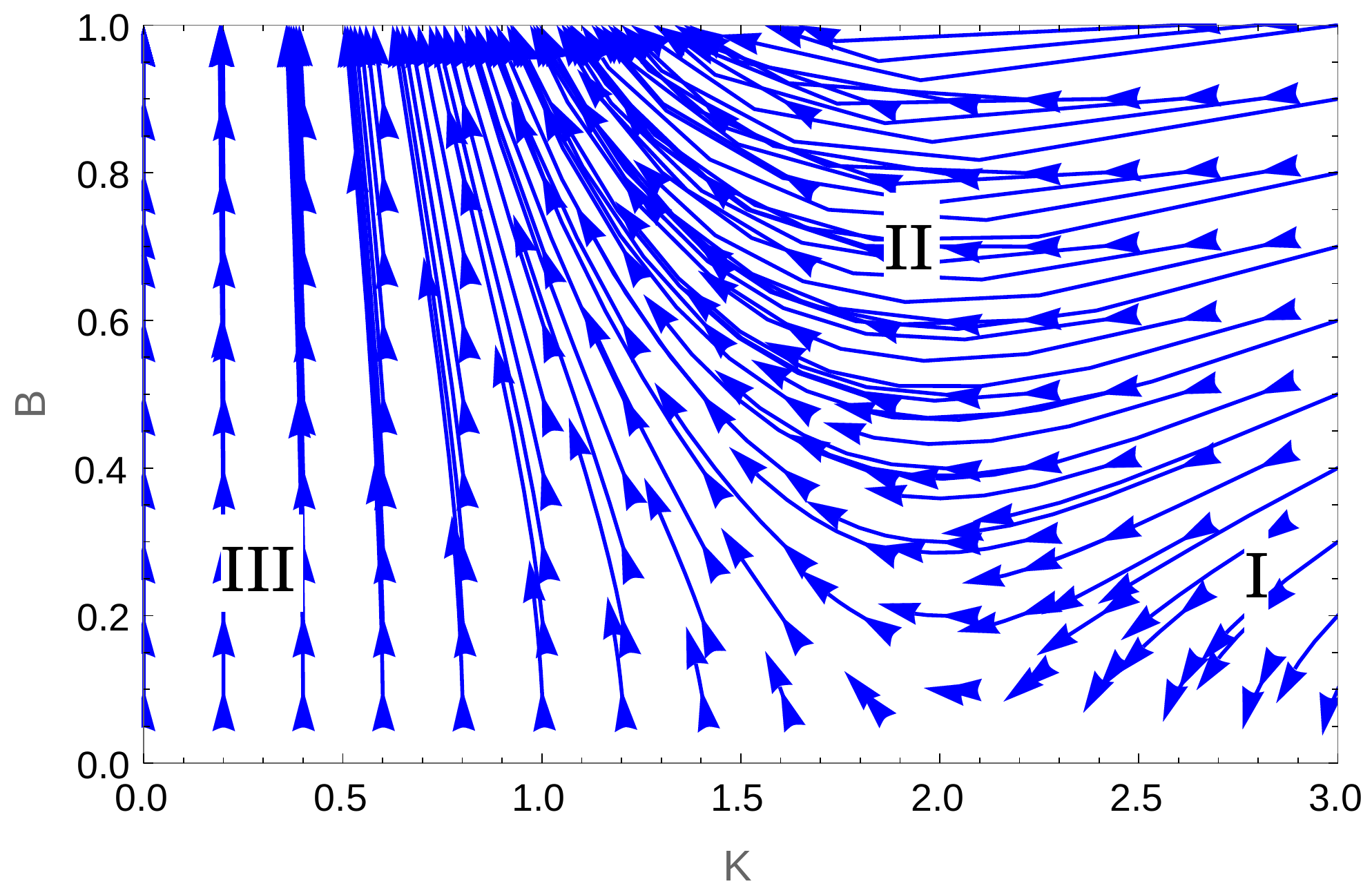}\label{RG2}}\\
						\caption{(a) RG flow for $\Delta$ with $K=\left( \frac{1}{2(y_{||}+1)}\right) $, 
					    (b) RG flow for $B$ with $K=\left( y_{||}+2\right) $, 
					    Arrow indicates the 
						direction of the RG flow.} 
					\end{figure} 
We define the family of hyperbola parameterized by $\alpha$,
\begin{equation}  y_{||}^2-\Delta^2=\alpha.\end{equation} 
Thus we have,
\begin{equation}
\begin{aligned}
\dfrac{d[ y_{||}^2-\Delta^2]}{dl}&= \left[ y_{||}\dfrac{dy_{||}}{dl} + y_{||}\dfrac{dy_{||}}{dl} \right] - 
\left[ \Delta \dfrac{d\Delta}{dl} + \Delta \dfrac{d\Delta}{dl} \right],\\
&= 2 y_{||}(-\Delta^2)-2\Delta(-y_{||}\Delta) = 0.
\end{aligned}
\end{equation}
Now we explain the different regime of the RG flow diagram.
We distinguish three different regimes based on the value of $\alpha$.
In fig.\ref{RG1}, we can define three regions, region I 
(weak coupling), region II (crossover) and region III (strong coupling). 
We follow ref.\cite{mudry2014lecture} during the explanation.\\

\noindent(1) When $\alpha>0$, parameterized hyperbolic equation is,\\

\;\;$y_{||}=(\pm)\sqrt{\alpha}\frac{1+\kappa^2}{1-\kappa^2}$, 
\;\;\;\;\;\; $\Delta=\sqrt{\alpha}\frac{2\kappa}{1-\kappa^2}$, 
\;\;\;\;\;\; $0\le \kappa < 1$\;\;
\begin{equation}
\frac{d\kappa}{dl}=(\mp)\sqrt{\alpha}\kappa.
\end{equation}
This is the RG equation for parameter $\kappa$ 
and the solution to this is,
$$ \int_{\kappa(l_0)}^{\kappa(l)} \frac{1}{\kappa} ds 
= \int_{0}^{l}(\mp)\sqrt{\alpha}dl,$$
$$ \ln\left( \frac{\kappa(l)}{\kappa(l_0)} \right) 
= (\mp) \sqrt{\alpha} l,$$
\begin{equation}
\kappa(l)=\kappa(l_0)e^{(\mp) \sqrt{\alpha} l}.
\end{equation} 
(1.a) When $\alpha>0$ and $y_{||}>0$ $\left( K<\frac{1}{2}\right) $,
\begin{equation}
\kappa(l)=\kappa(l_0)e^{-\sqrt{\alpha} l}.
\end{equation} 
This shows the $\kappa(l)$ decreases with length scale showing the weak coupling phase. The region I is the weak coupling phase. 
In this phase, there is no gaped excitation, 
i.e, region I is in the gapless helical Luttinger liquid phase where the 
sine-Gordon coupling term is irrelevant. In this phase, there is no evidence of Majorana
fermion mode, i.e, system is in the non-topological state.\\
(1.b) When $\alpha>0$ and $y_{||}<0$ $ \left( K>\frac{1}{2}\right) $,
\begin{equation}
\kappa(l)=\kappa(l_0)e^{+\sqrt{\alpha} l}.
\end{equation} 
It is very clear from the above equation that $\kappa(l)$ increases with length scale. As a consequence of it, RG flow lines flowing off to the deep massive phase. The region III is the deep massive phase, i.e, the sine-Gordon coupling term is relevant, and the 
RG flows flowing off to the strong coupling regime away from the Gaussian fixed line.\\

\noindent(2) Now we do the analysis for crossover phase (region II). When $\alpha<0$, parameterized hyperbolic equation is,\\

$y_{||}=\sqrt{|\alpha|}\frac{2\kappa}{1-\kappa^2}$, 
\; $\Delta=\sqrt{|\alpha|}\frac{1+\kappa^2}{1-\kappa^2}$, \; $-1 < \kappa < 1$.
\begin{equation}
\frac{d\kappa}{dl}=-\frac{\sqrt{|\alpha|}}{2}(1+\kappa^2).
\end{equation} 
This is the RG equation for the parameter $\kappa$ 
and the solution to this is,
$$ \int_{\kappa(l_0)}^{\kappa(l)} \frac{1}{1+\kappa^2} ds 
= -\frac{\sqrt{|\alpha|}}{2} \int_{l_0}^{l}dl,$$
\begin{equation}
\tan^{-1}(\kappa(l))-\tan^{-1}(\kappa(l_0))= -\frac{\sqrt{|\alpha|}}{2}(l-l_0).
\end{equation} 
The region II is the crossover regime. 
One observes
the crossover from the weak coupling phase 
to the strong coupling region. 
During this phase, the system transits from gapless 
phase to the proximity
induced superconducting gaped phase, i.e., $\Delta \neq 0 $. 
This is the topological superconducting phase, where the
system has the Majorana fermion mode. For this situation, 
system transits
from the non-topological state to the topological state. 
In practical reality, for this regime of parameter space 
the quantum spin Hall insulator will be in the topological 
state of matter with Majorana edge mode.\\ 
The difference between the region II and region III is, 
in region II, the field never reaches to free scalar 
field, i.e., ignoring the potential either at the long 
distance or at the short distance physics.\\\\
\textbf{ Quantum BKT equations and results 
for Hamiltonian $H_2$}\\
Here we present the quantum BKT equations 
of the Hamiltonian $H_2$ and observe there 
is no topological phase,
\begin{equation}
H_2 =\frac{v}{2} \int \left[ \frac{1}{K}(\partial_x \phi(x))^2 
+ K(\partial_x \theta(x))^2\right] dx+ \frac{B}{\pi} \int  \cos(\sqrt{4\pi}\phi(x)) dx.
\end{equation} 
Following the same procedure of appendix A, 
we obtain the RG equations for the Hamiltonian $H_2$ as,
\begin{equation}
\dfrac{dB}{dl}= B(2-K), \;\;\;\;\;\;\;\;
\frac{d K}{dl} = - B^2 K^2. \label{RGb}
\end{equation}
We do the transformation, $y_{||} = (K-2) $
and obtain another set of RG equations in the standard form of BKT equation,
\begin{equation}
\frac{dB}{dl} = -y_{||} B, \;\;\; \frac{dy_{||}}{dl} = -B^2. \label{RGb1}
\end{equation}
The structure of the second BKT equations (eq.\ref{RGb1}) 
is same as that of the first one (eq.\ref{RGdelta1}). 
Therefore the analysis of this 
equation is the same as that of the first one. The only 
difference being, there is no evidence Majorana  fermion
mode in this phase. In fig.\ref{RG2}, the system shows 
only the Ising-ferromagnetic phase (region III). In practical reality, for this
regime of parameter space the quantum spin Hall 
insulator will be in the non-topological phase. 
From this study, we obtain two types of BKT equations, 
but there is no Majorana-Ising transition. We observe 
Majorana-Ising transition, if we consider total RG equations
of Hamiltonian $H$ (eq.\ref{model1}) \cite{Sarkar-2016} in which case the quantum BKT behavior will be absent.\\

\textbf{\large Exact solution of the RG equations}\\
\noindent 
The model Hamiltonian gives two set of RG
equation which yield the quantum BKT equation. 
Here we derive analytical solution for these two RG 
equations by solving them directly. Let us consider the first set of RG equation,
\begin{equation} 
\frac{d \Delta}{dl} = \left( 2 -\frac{1}{K}\right)  \Delta, \;\;\;\;\;\;\;\; \frac{dK}{dl} = {\Delta^2}.\end{equation} 
\noindent We use the above two equation to derive the 
phase relation between the two parameters $\Delta$ and K.
\begin{equation}  
\frac{d\Delta}{dK} = \frac{\left( 2-\frac{1}{K}\right)\Delta }{\Delta^2}
 = \frac{\left(
2-\frac{1}{K}\right)}{\Delta}\end{equation} 
On integration, we get C as,
\begin{equation}   C = \frac{\Delta^2}{2} - 2K +\ln K  \end{equation} 
\noindent We can calculate $C$ from the initial values $\Delta_0$, $K_0$. 
\begin{equation}  \frac{\Delta^2}{2} = \frac{\Delta_0^2}{2} 
+ 2(K-K_0)-\ln \left(\frac{K}{K_0}\right) \end{equation}  
\begin{equation}  \Delta = \sqrt{\Delta_0^2 + 4(K-K_0)
-2\ln \left(\frac{K}{K_0}\right)}.\end{equation}
Here $\Delta_0 $ and $K_0 $ are the initial value of $K$ and $\Delta$.
\begin{figure}[h]
\subfloat[]{\includegraphics[scale=0.4]{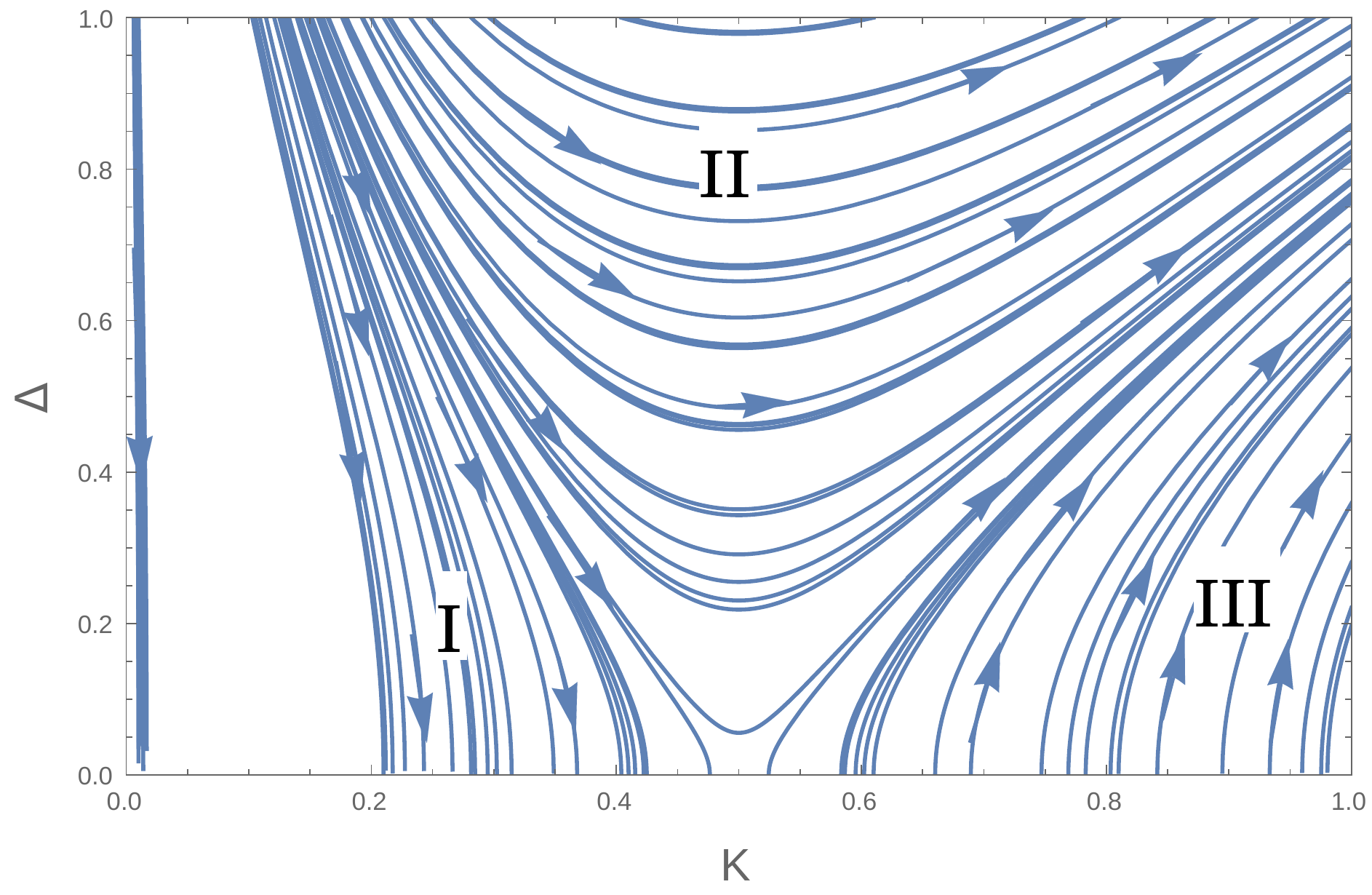}\label{figure1a}}
\;\;\;\subfloat[]{\includegraphics[scale=0.4]{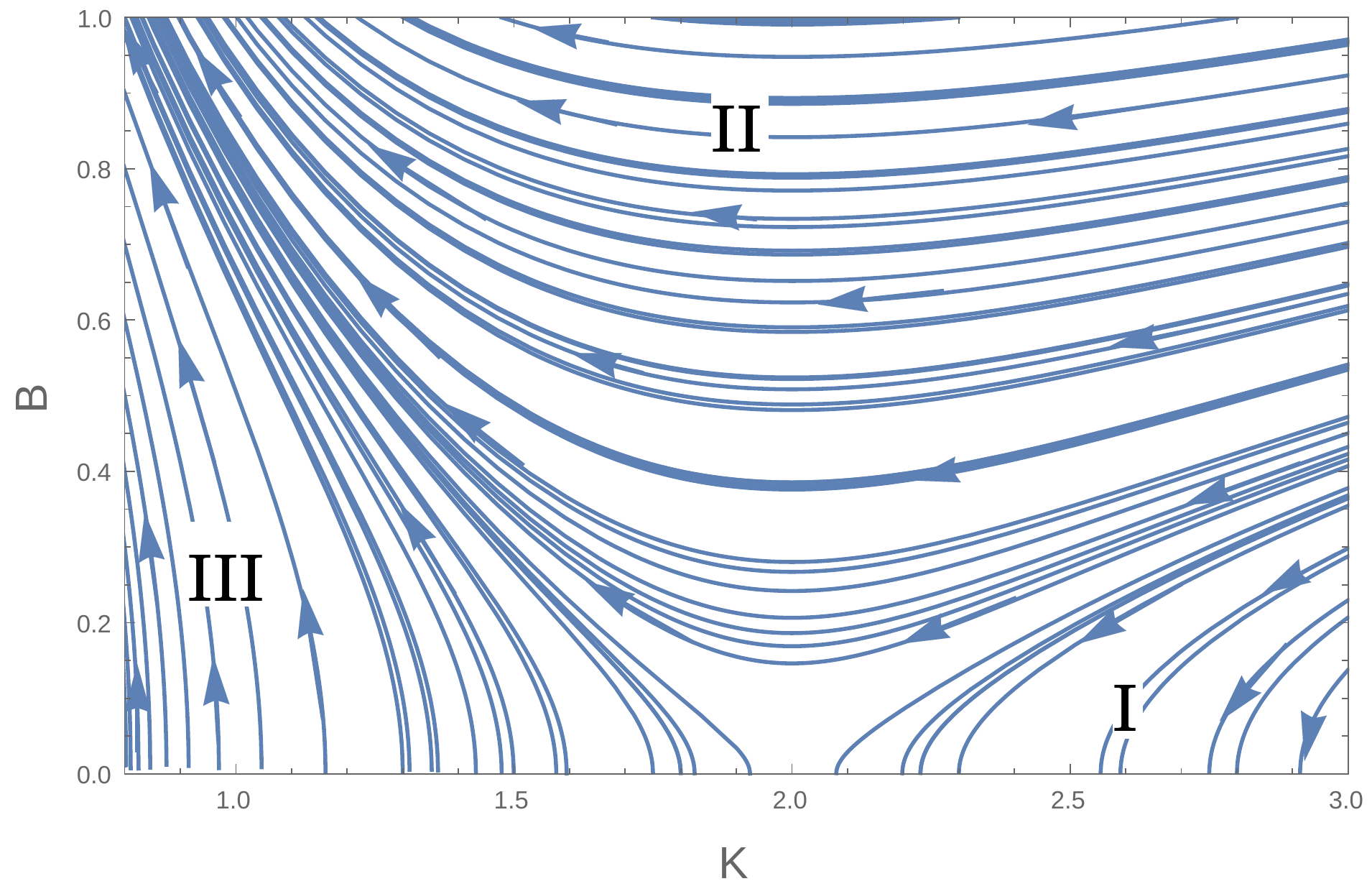}\label{figure1b}}\\
\caption{(a) The curve is plotted for $\Delta$ with $K$.  
Analytical relation of $K$ and $y_{||}$ is $K=\left( \frac{1}{2(y_{||}+1)}\right) $. 
(b) The curve is plotted for $B$ with $K$.  
Analytical relation of $K$ and $y_{||}$ is $K=\left( y_{||}+2\right) $. 
The arrow indicates the direction of the RG flow.}
\end{figure} 
In fig.\ref{figure1a}, 
one can observe three regions, 
region I corresponds weak coupling gapless helical Luttinger liquid phase, where the
RG flow lines flowing off to the weak coupling phase. It is a Gaussian fixed line. 
Region II is the crossover regime 
from weak coupling to strong coupling phase. 
Region III corresponds to strong coupling deep massive phase, 
away from Gaussian fixed line,
where RG flow lines are flowing off
from the weak coupling phase to the strong coupling phase, i.e., the system
flowing off from gapless helical LL phase to topological superconducting phase. 
In region III, one can observe the asymptotic nature of the system. 
In region III flow lines indicate the increase in the length scale, 
one can observe the coupling constant increases as the 
length scale increases and vice versa.\\
Now let us consider the second set of RG equation,
\begin{equation}
\frac{d{B}}{dl} = (2 - K) {B}, \;\;\;\;\;\;\;\;
\frac{dK}{dl} = - K^2  {B}^2 .
\end{equation}
\noindent Now we define a quantity $C$ as,
\begin{equation}
\frac{dB}{dK} = \frac{(2-K)B}{-K^2B^2} = \frac{(2-K)}{-K^2B}
\end{equation}
\noindent On integration, we get
$ C = \frac{B^2}{2} - \frac{2}{K} -\ln K $\\
	
\noindent We can calculate $C$ from the initial values $B_0$, $K_0$.
	
$$ \frac{B^2}{2} = \frac{B_0^2}{2} + 2(\frac{1}{K}-\frac{1}{K_0})+\ln \left(\frac{K}{K_0}\right) $$ 
	
\begin{equation}
B =\sqrt{B_0^2 + 4(\frac{1}{K}-\frac{1}{K_0})+2\ln \left(\frac{K}{K_0}\right)}
\end{equation}
In fig.\ref{figure1b}, 
one can observe three regions, 
region I corresponds weak coupling gapless helical Luttinger liquid phase, where the
RG flow lines flowing off to the weak coupling phase. It is a Gaussian fixed line. 
Region II is the crossover regime 
from weak coupling to strong coupling phase, initially RG flow lines flowing off
to the weak coupling regime but due to the presence of $\frac{dK}{dl}$ in the RG equation, the RG flow lines flowing off to the strong coupling phase,  
which is the
Ising-ferromagnetic phase. 
Region III, corresponds to strong coupling deep massive phase, 
away from Gaussian fixed line, 
where RG flow lines flowing off
from the weak coupling phase to the strong coupling phase, i.e., the system
flowing off from gapless helical LL phase to Ising-ferromagnetic phase, which is non-topological
in character. 
In region III, one can observe the asymptotic nature of the system. 
In region III flow lines indicate that the 
coupling constant increases as the 
length scale.\\

{\bf  Effect of chemical potential on RG flow lines :} \\
We have two model Hamiltonian to study the quantum BKT. One is for the $\phi$ field and the other is for the $\theta$ field. The chemical potential ($\mu$) is related with the $\phi$ field. The effect of $\mu$ will be different for the two quantum BKT transitions. At first we study the effect of $\mu$ for the $\phi$ field.\\
\noindent\textbf{Effect on the Hamiltonian $H_2$ :}
Here we study how the chemical potential
affect the Hamiltonian $H_2$. We can absorb the $\partial_x \phi$ term into the quadratic Hamiltonian by the transformation of $\phi$ field, $\phi \rightarrow \phi - \frac{K \mu}{\sqrt{\pi}} x$. This gives the spatially oscillating term which modify the cosine term as, $\frac{B}{\pi} cos (2 \sqrt{\pi} \phi + \delta_1 \phi (x))$ indicating commensurate to incommensurate transition. However one can write another RG equation to study the effect of $\mu $ ,i.e, $ \frac{d \delta_1}{dl} = \delta_1$ \cite{giamarchi2003quantum,Sarkar-2016, Lutchyn}. Under the condition $\delta_1(0)a<<B(0)^{\frac{1}{(2-K)}}$, if $B(l)$ reaches to strong coupling phase before $\delta_1(l)a \rightarrow 1$ then the system is in Ising-ferromagnetic phase. Therefore, it is clear from the above study based on the RG equation of $\delta_1 $ that the 
different quantum phases of this system dominates in presence of chemical
potential in the different regime of the interaction space.\\
\textbf{Effect on the Hamiltonian $H_1$ :} 
We consider the Hamiltonian $H_1$ in the presence of chemical potential, which we call $H_3$, and find the effect of chemical potential in RG flow diagrams. Here $\phi$ and $\theta$ fields are dual to each other (i.e. minima of the sine-Gordon coupling term for $\phi$ and $\theta$ are different), and the relation between these two fields is
\begin{equation*}
[\phi(x),\partial_{x^{\prime}} \theta(x^{\prime})=i\delta(x-x^{\prime})].
\end{equation*}
Therefore we are not allowed to absorb the $\phi$ field in the sine-Gordon coupling term of $\theta$.
Now the Hamiltonian can be written as
\begin{equation} H_3 = \frac{v}{2} \int \left[ \frac{1}{K}(\partial_x \phi(x))^2 + K(\partial_x \theta(x))^2\right] dx  -\frac{\Delta}{\pi} \int \cos(\sqrt{4\pi}\theta(x)) dx  - \frac{\mu}{\sqrt{\pi}}\int \partial_x \phi(x) dx.  \label{withmu}
\end{equation} 
The quantum BKT equations of $H_3$ is given by (for detailed derivation see appendix B),
\begin{equation}
\frac{d\Delta}{dl} = \left[ 2 - \frac{1}{K}\left( 1 + \frac{\mu}{v\pi}\right) \right]  \Delta, \;\;\;\;\;\;\;
\frac{dK}{dl} = \Delta^2 .  \label{RGdeltamu}
\end{equation}
\begin{figure}
\begin{center}
\includegraphics[scale=0.45]{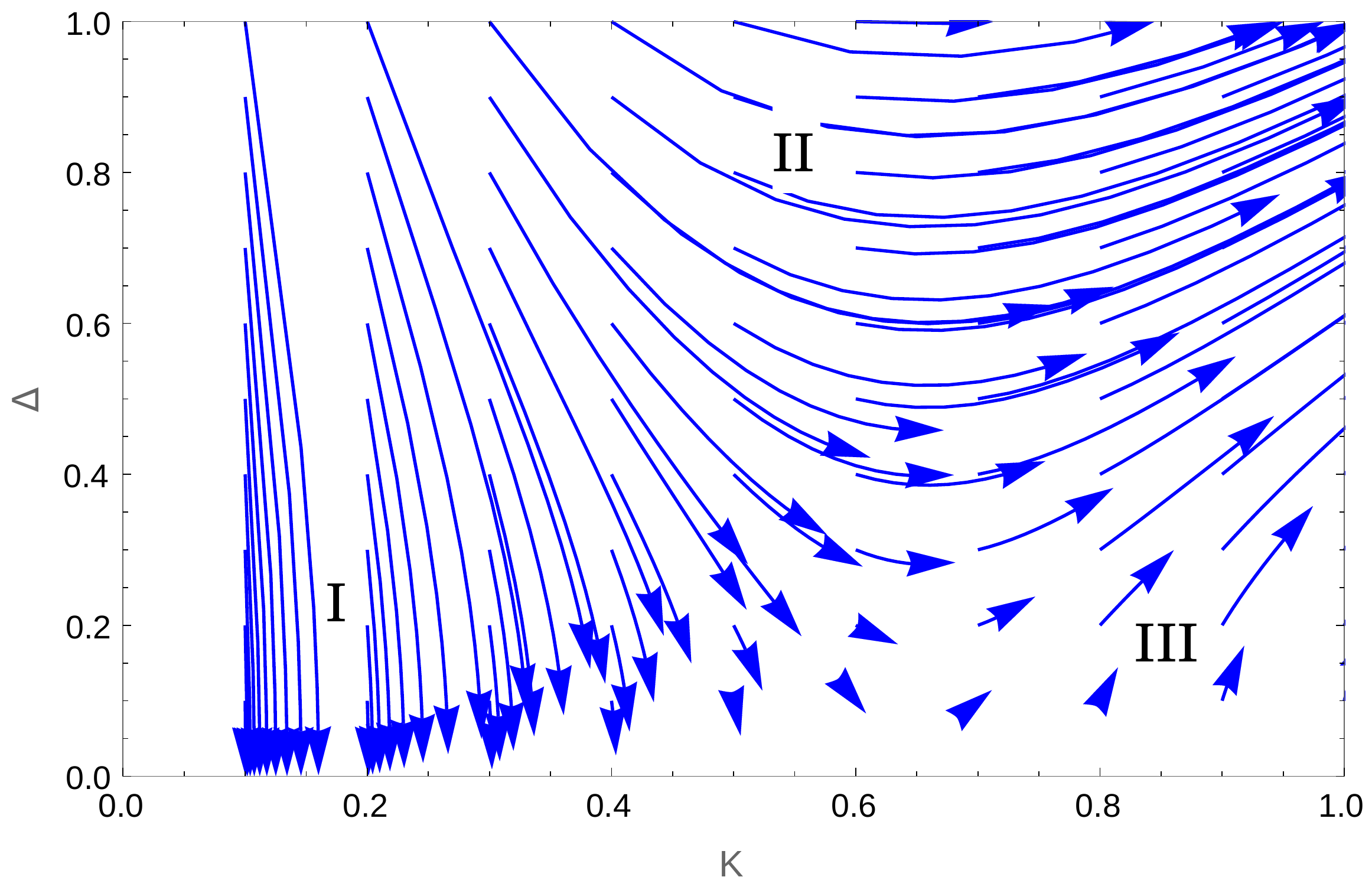}
\caption{Renormalization group flow for $\Delta$ with $K$ for both the sets of RG equations for finite $\mu$ ($\mu=1$), arrow indicates the direction of the RG flow.}
\label{RG3}
\end{center}
\end{figure}
It is clear from the eq.\ref{RGdeltamu} that in the presence of finite $\mu$, the analytical form of the equation is the same as that of $\mu=0$, but with a modification of a factor. In fig.\ref{RG3}, we present the results for finite $\mu$ ($\mu=1$), the behavior of the RG equation remain same. Therefore it reveals from our study that the existence of Majorana fermion mode does not disappear for finite $\mu$.\\\\ 
{\bf Summary of the new and important results of the present quantum BKT study
:}\\
We have obtained three quantum phases either topological or non-topological in character from the study of  quantum BKT RG equations. 
One is  topological, i.e., topological superconducting phase, another one is the
Ising-ferromagnetic phase which is non-topological and finally we have obtained gapless
helical Luttinger liquid phase, which is also non-topological in character. The region I is the helical Luttinger liquid phase where the RG flow lines flowing off the weak coupling phase.
The system reaches the strong coupling phase when RG flow lines flows from region II to region III, where the sine-Gordon coupling term become relevant, 
it is topological superconducting phase or Ising-ferromagnetic phase.
In region-III, field theory is asymptotically free, i.e, the system reduced to the
scalar field theory by ignoring the sine-Gordon coupling term at short distance. This short distance physics of the region III, does not appear in region II.\\

\begin{figure}
\begin{center}
\includegraphics[scale=0.3]{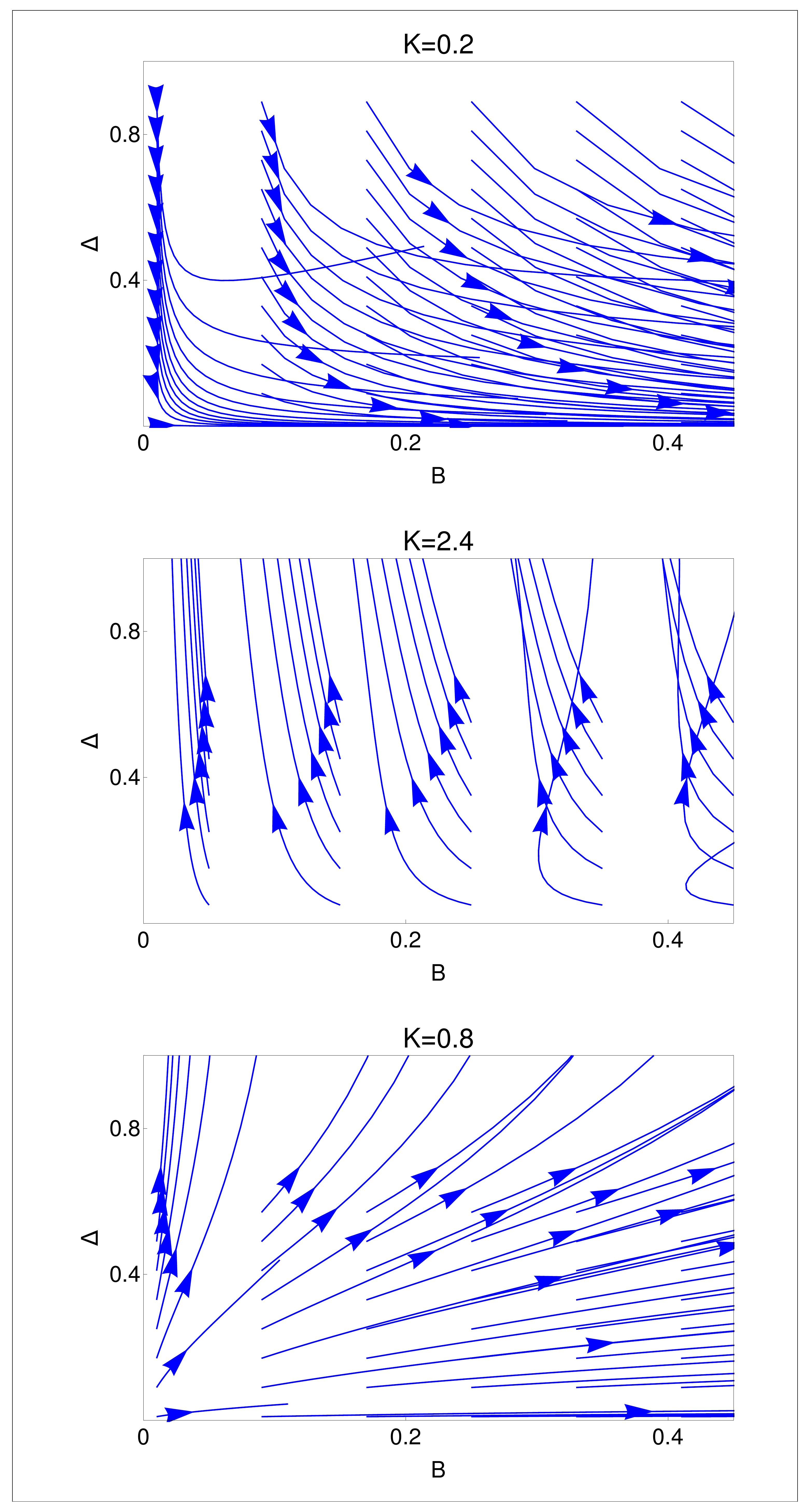}
 \caption{Phase diagram showing, upper panel : Ising-ferromagnetic phase ($K=0.2$); middle panel : Majorana phase ($K=2.4$); lower panel : Majorana-Ising transition ($K=0.8$).} \label{delta-B}
\end{center}
\end{figure}
\noindent \textbf{A comparison between the results of quantum BKT and with the results of total Hamiltonian:}\\
In this section we compare the results which we have obtained from the study of two quantum BKT Hamiltonians and the total Hamiltonian of the system (eq.\ref{model2}). The total Hamiltonian of the system has already been studied in different context \cite{giamarchi2003quantum}. Very recently, the Hamiltonian in eq.\ref{model1} has also been studied in the context of topological states of matter in ref.\cite{Altland-2011},\cite{Sarkar-2016} and \cite{Sudip-kumar}.   
We consider the Hamiltonian eq.\ref{model2} (without $\mu$), which yields the RG equations for $\Delta$, $B$ and $K$.
 \begin{equation}
 	\begin{split}
 	H  = \frac{v}{2}\int  \left[ \frac{1}{K} ( {({\partial_x \phi(x) })}^2 + 
 	K {({\partial_x  \theta(x) })}^2  )\right] dx
 	+\frac{B}{\pi} \int cos(\sqrt{4 \pi} \phi(x)) dx - \frac{\Delta}{\pi}\int cos(\sqrt{4 \pi \theta(x)}) dx  
 	\end{split} \label{model11}
 	\end{equation}
The RG equations can be derived as,
\begin{align}
\frac{dB}{dl} &= \left( 2-K \right)B, \\ \frac{d\Delta}{dl} &= \left(2 - \frac{1}{K}\right)   \Delta, \\
\frac{dK}{dl} &= \frac{1}{2\pi^2}\left( \Delta^2 - B^2K^2 \right) .
\end{align}
These RG equation consist to two sine-Gordon coupling term, one is the $\Delta$, which induce the topological superconducting phase and the other is $B$, which induce the Ising-ferromagnetic phase in the system. 
$K$ and $\mu$ are the parameters of the model Hamiltonian. In the present section we present the results of the whole RG equation for the different values of $K$. 
$K<1$ and $K>1$ 
characterizes the repulsive and attractive interactions 
respectively, where as $K=1$ characterizes non-interacting case \cite{giamarchi2003quantum}.\\
Fig. \ref{delta-B} consist of three panels for different values of Luttinger liquid parameter to show the existence of different quantum phases either topological or non-topological in character and also the transition between them. The upper panel ($K=0.2$) of the figure present the existence of Ising-ferromagnetic phase. This is because the RG flow lines flowing off to the weak coupling phase for the coupling $\Delta$, but the coupling $B$ increases. The middle panel is for $K=2.4$. We observe that the system is in the topological state i.e, the coupling $\Delta$ increases to the strong coupling phase but there is no increase of coupling $B$. The lower panel is for $K=0.8$. We observe Majorana-Ising transition in the system. RG flow lines increases for large initial values of $\Delta$ than $B$, thus the system is driven to the topological state. Otherwise the system is in the Ising-ferromagnetic phase. Therefore we conclude that from the RG flow lines that, there is no evidence of helical LL phase for this study. But for the quantum BKT study we have not observed any Majorana-Ising transition. \\
In quantum BKT instead of two sine-Gordon coupling term, there is only one sine-Gordon coupling term for each Hamiltonian. The sine-Gordon coupling term gives the confining potential and the quadratic part of the Hamiltonian (eq.\ref{model11}) gives the kinetic energy contribution. Therefore the compitition between these kinetic energy (quadratic fluctuation) and sine-Gordon coupling terms finally gives winning phase of the system, instead of compitition between the two sine-Gordon coupling term of the total RG equation. Therefore in this quantum BKT there is no Majorana-Ising transition. At the same time in the quantum BKT we present the results for RG flow diagram for a single coupling constant with $K$.\\
\textbf{ Conclusion :}
We have studied quantum BKT transition for the one-dimensional interacting edge mode of helical
liquid of topological insulator 
and have also found two quantum BKT transition for different
physical situations. We  have shown the existence  of topological superconducting phase, gapped  Ising-ferromagnetic phase and gapless helical Luttinger  liquid phase for
this system  through RG flow  diagrams. We have found the exact
solution for quantum BKT transition for the helical edge state system which appears 
in the quantum spin Hall system. For finite chemical potential, we also observe
the presence of commensurate to incommensurate transition.  We have not found any direct Majorana-Ising transition in quantum BKT transitions.\\

\textbf{ Acknowledgment :}
S.S would like  to acknowledge
DST (EMR/2017/000898)  for the support. R.K.R and R.S would like  to acknowledge PPISR, RRI library for  the books
and  journals and ICTS
Lectures/seminars/workshops/conferences/discussion     meetings     of
different aspects of physics.


%

\bibliography{references}

\textbf{\large Appendix}
\appendix
\textbf{\large A) Derivation of Quantum BKT equations for $H_1$}\\
The Hamiltonian $H_1$ is given by,
\begin{equation} 
H_1 = \frac{v}{2}\int \left[ \frac{1}{K} (\partial_x \phi(x))^2 + K(\partial_x  \theta(x))^2\right] dx - \frac{\Delta}{\pi}
 \int \cos(\sqrt{4\pi}\theta(x)) dx .
\end{equation} 
In  the Bosonized  model
Hamiltonian  $H_1$,  we  rescale  the  fields  as,  $\phi  \rightarrow
\phi^{\prime}=\phi/\sqrt{K}$       and       $\theta       \rightarrow
\theta^{\prime}=\sqrt{K}\theta$.   Thus  the  quadratic  part  of  the
Hamiltonian will be,
\begin{equation} H^{\prime}_q = \frac{v}{2}[(\partial_x \phi^{\prime})^2 + (\partial_
    x \theta^{\prime})^2] .\end{equation} The Hamilton's equations for
the    cannonically   conjugate    fields   ($\phi^{\prime}$    an   d
$\theta^{\prime}$) are,
\begin{equation} \partial_x \theta^{\prime} = -\frac{1}{v}\partial_t \phi^{\prime} \;
  \;\;,
\partial_x        \phi^{\prime}        =        -\frac{1}{v}\partial_t
\theta^{\prime}.
\end{equation} 
Thus the Lagrangian in terms of $\theta^{\prime}$ field is given by,
\begin{equation}
\begin{aligned}
\mathcal{L}_0&=            \Pi_{\theta^{\prime}}            \partial_t
\theta^{\prime}-H_q^{\prime},\\   &=    \left(   \frac{1}{v}\partial_t
\theta^{\prime}        \right)       \partial_t        \theta^{\prime}
-\frac{v}{2}(\partial_x   \theta^{\prime})^2-   \frac{v}{2}(\partial_x
\phi^{\prime})^2,\\          &=          \frac{1}{2}[v^{-1}(\partial_t
  \theta^{\prime})^2-v(\partial_x \theta^{\prime})^2].
\end{aligned}
\end{equation}
The Lagrangian rewritten in the imaginary time ($\tau=it$) as,
\begin{equation}
\mathcal{L}_0= -\frac{1}{4}[v^{-1}(\partial_{\tau}\theta^{\prime})^2 +
  v(\partial_{x}\theta^{\prime})^2 ],
\end{equation} 
The     Lagrangian     for     interaction     term     will     have,
$\mathcal{L}_{\Delta}=-H_{\Delta}$, where,
\begin{equation}
\mathcal{L}_{\Delta}         =         \left(\frac{\Delta}{\pi}\right)
\cos(\sqrt{4\pi}\theta(x)) .
\end{equation} 
The Euclidean action  can be written as, $S_E= -\int  dr \mathcal{L} =
-\int    dr    (\mathcal{L}_0    +    \mathcal{L}_{\Delta})$,    where
$r=(\tau,x)$.  
Now  we  write  the  partition  function  in  terms  of
Euclidean action,
\begin{equation}
\mathcal{Z}=   \int  \mathcal{D}\theta^{\prime}   e^{\left[  \int   d\tau dx
    \left(-\frac{1}{4}v^{-1}(\partial_{\tau}\theta^{\prime})^2
    -\frac{1}{4}  v(\partial_{x}\theta^{\prime})^2\right)  -  \int  d\tau dx
   \mathcal{L}_{\Delta}(\theta^{\prime}) \right]}.\\
\end{equation}
We write partition fucntion in a local form using space independent fields ($\tilde{\theta}^{\prime}$), which describe the system at the point contact, i.e. at $x=0$ \cite{Andres}. To perform this we integrate the fields everywhere except $x=0$. The partition function is
\begin{equation}
\mathcal{Z}=   \int  \mathcal{D}\theta^{\prime}  \mathcal{D}\tilde{\theta}^{\prime}
 e^{-S_E(\theta^{\prime})} \delta(\tilde{\theta}^{\prime}(\tau) - \theta^{\prime}(\tau,0),\\ \label{parti}
\end{equation}
where $\delta(\tilde{\theta}^{\prime}(\tau) - \theta^{\prime}(\tau,0) = \frac{1}{2\pi} \int dk_{\theta^{\prime}}(\tau) e^{ik_{\theta^{\prime}}(\tilde{\theta}^{\prime}-\theta
^{\prime}(\tau,0))}$. We first solve for the integral
\begin{equation}
I=\int   d\tau dx
    \left(v^{-1}(\partial_{\tau}\theta^{\prime})^2
    +  v(\partial_{x}\theta^{\prime})^2\right)
\end{equation}
One can rewrite this integral as Fourier sums, which yields
\begin{dmath}
I= \int d\tau dx \left[
v^{-1}\frac{1}{\beta L} \sum_{q,\omega_n} \left(-i \omega_n \theta^{\prime}_{q,\omega_n} \right) e^{i(qx-\omega_n \tau)} 
\times 
\frac{1}{\beta L} \sum_{q^{\prime},\omega^{\prime}_n} \left(i \omega^{\prime}_n \theta^{\prime *}_{q^{\prime},\omega^{\prime}_n} \right) e^{i(q^{\prime}x-\omega^{\prime}_n \tau)} \\
+
v \frac{1}{\beta L} \sum_{q,\omega_n} \left(i q \theta^{\prime}_{q,\omega_n} \right) e^{i(qx-\omega_n \tau)} 
\times 
\frac{1}{\beta L} \sum_{q^{\prime},\omega^{\prime}_n} \left(-i q^{\prime}_n \theta^{\prime *}_{q^{\prime},\omega^{\prime}_n} \right) e^{i(q^{\prime}x-\omega^{\prime}_n \tau)} 
\right] \\
= \frac{1}{\beta L} \sum_{q,q^{\prime}, \omega_n,\omega^{\prime}_n} \left( vqq^{\prime} + v^{-1}\omega_n \omega^{\prime}_n\right) \theta^{\prime }_{q,\omega_n}\theta^{\prime *}_{q^{\prime},\omega^{\prime}_n} \delta_{q,q^{\prime}}\delta_{\omega_n, \omega^{\prime}_n}\\
= \frac{1}{\beta L} \sum_{q,\omega_n} \left( vq^2 + v^{-1} \omega_n^2\right) |\theta^{\prime}|^2. 
\end{dmath}
Similarly one can solve other two integrals in the exponential of eq.\ref{parti}.
\begin{equation}
i \int d\tau k_{\theta^{\prime}}\tilde{\theta}^{\prime}(\tau) = \frac{1}{\beta} \sum_{\omega_n} k(\omega_n) \tilde{\theta}^{\prime}(-\omega_n).
\end{equation}
\begin{equation}
-i \int d\tau k_{\theta^{\prime}} \theta^{\prime}(\tau,0)= -\frac{i}{\beta L} \sum_{q,\omega_n} k (-\omega_n) \theta^{\prime}_{q,\omega_n}.
\end{equation}
Partition function can now be rewritten by substituting the above integrals as
\begin{equation}
\mathcal{Z}= \int \mathcal{D}\theta  \mathcal{D}k_{\theta}
e^{-\frac{1}{4\beta L}\sum_{q,\omega_n}  \left[ \left( \frac{K}{v}\omega_n^2 + Kvq^2\right) |\theta|^2  -4ik_{\theta} (-\omega_n) \theta(q,\omega_n) \right] 
+ \frac{i}{\beta} \sum_{\omega_n} k_{\theta} (\omega_n) \tilde{\theta}(-\omega_n) 
- \int d\tau dx \mathcal{L}_{\Delta}(\tilde{\theta})},
\end{equation}
here we have transformed $\theta^{\prime }$ fields back to $\theta$ fields. Now we perform Gaussian integration over $\theta$ field. 
\begin{equation}
\mathcal{Z}\propto \int \mathcal{D}\tilde{\theta} \mathcal{D}k_{\theta} e^{\frac{1}{\beta L} \sum_{q,\omega_n} k_{\theta} (-\omega_n)  \left( \frac{K}{v}\omega_n^2 + Kvq^2\right)^{-1} k_{\theta} (\omega_n) + \frac{i}{\beta} \sum_{\omega_n} k_{\theta} (\omega_n) \tilde{\theta}(-\omega_n) 
- \int d\tau dx \mathcal{L}_{\Delta}(\tilde{\theta})}.
\end{equation}
Taking the q-sums to the continuum limit and performing the resulting integral
\begin{align}
\mathcal{Z} &\propto \int \mathcal{D}\tilde{\theta} \mathcal{D}k_{\theta} e^{-\frac{1}{\beta}\sum_{\omega_n} \left[ k_{\theta(-\omega_n) k_{\theta}(\omega_n)} \frac{1}{K} \int \frac{dq}{2\pi} \left(\frac{1}{v} \omega_n^2 + vq^2 \right)^{-1} \right] 
+ \frac{i}{\beta} \sum_{\omega_n} k_{\theta} (\omega_n) \tilde{\theta}(-\omega_n) 
- \int d\tau dx \mathcal{L}_{\Delta}(\tilde{\theta})}\\
&\propto \int \mathcal{D}\tilde{\theta} \mathcal{D}k_{\theta} e^{-\frac{1}{\beta}\sum_{\omega_n} \left[ k_{\theta(-\omega_n) k_{\theta}(\omega_n)} \frac{1}{K} \left( \frac{1}{2|\omega_n|} \right) \right] 
+ \frac{i}{\beta} \sum_{\omega_n} k_{\theta} (\omega_n) \tilde{\theta}(-\omega_n) 
- \int d\tau dx \mathcal{L}_{\Delta}(\tilde{\theta})}
\end{align}
Finally performing the Gaussian integration over $k_{\theta}$ gives
\begin{equation}
\mathcal{Z} = \int \mathcal{D}\tilde{\theta} e^{-\frac{1}{2\beta} \sum_{\omega_n} \left( K |\omega_n| |\tilde{\theta}|^2 \right)  - \int d\tau dx \mathcal{L}_{\Delta}(\tilde{\theta})}
\end{equation}
In the continuum limit of $\omega_n$ we have
\begin{equation}
\mathcal{Z} = \int \mathcal{D}\tilde{\theta} e^{-\int\limits_{-\Lambda}^{\Lambda}\frac{d\omega}{2\pi}|\omega| \frac{K |\tilde{\theta}(\omega)|^2}{2} - \int d\tau dx \mathcal{L}_{\Delta}(\tilde{\theta})}
\end{equation}
Final form of the partition function can be written as,
\begin{equation}
\mathcal{Z}=         \int          \mathcal{D}\theta         e^{\left[
    -\int_{-\Lambda}^{\Lambda}            \frac{d\omega}{2\pi}|\omega|
    \frac{K|\theta(\omega)|^2}{2}          -          \int          dr
    \mathcal{L}_{\Delta}(\theta)\right]}.
\end{equation}
We now  separate the slow and  fast fields and integrate  out the fast
field       components.       The        field       $\theta$       is
$\theta(r)=\theta_s(r)+\theta_f(r)$ where,
\begin{equation}
\theta_s(r)=\int_{-\Lambda/b}^{\Lambda/b}         \frac{d\omega}{2\pi}
e^{-i\omega\tau}\theta(\omega)                        \;\;\;\;\&\;\;\;
\theta_f(r)=\int_{\Lambda/b<|\omega_n|<\Lambda}   \frac{d\omega}{2\pi}
e^{-i\omega\tau}\theta(\omega),
\end{equation}
here $r=(x,\tau)$. Now the partition function can be written as,
\begin{equation}
\begin{aligned}
\mathcal{Z}&=\int    \mathcal{D}\theta_s     \mathcal{D}\theta_f    e^
        {[-S_s(\theta_s)-S_f(\theta_f)-
            S_{\Delta}(\theta_s,\theta_f)]},\\         &=         \int
        \mathcal{D}\theta_s  e^ {[-S_s(\theta_s)]}  \left\langle e^{[-
            S_{\Delta}(\theta_s,\theta_f)]}\right\rangle_f,
\end{aligned}
\end{equation}
where      we     have      used     $\left\langle      F\right\rangle
_f=\int\mathcal{D}\theta_f   e^{-S_f(\theta_f)}F$.    We   write   the
effective action as,
\begin{equation}
e^{-S_{eff}(\theta_s)}=e^{-S_s(\theta_s)} \left\langle e^{-S_{\Delta}(\theta)} \right\rangle_f.
\end{equation}
Taking $\ln$ on both side gives,
\begin{equation}
S_{eff}(\theta_s)=        S_s(\theta_s)       -        \ln\left\langle
e^{-S_{\Delta}(\theta)} \right\rangle_f.
\end{equation}
By writing the cumulant expansion up to 2nd order, we have
\begin{equation}
S_{eff}(\theta_s)=                          S_s(\theta_s)+\left\langle
S_{\Delta}(\theta_s,\theta_f)\right\rangle  - \frac{1}{2}(\left\langle
S_{\Delta}^2(\theta_s,\theta_f)\right\rangle-\left\langle
S_{\Delta}(\theta_s,\theta_f)\right\rangle^2).
\end{equation}
Now  we   calculate  the   first  order   approximation  $\left\langle
S_{\Delta}(\theta_s,\theta_f)\right\rangle$,
\begin{equation}
\begin{aligned}
\left\langle        S_\Delta(\theta_s,\theta_f)\right\rangle        &=
\frac{\Delta}{\pi}\int\mathcal{D}\theta_f   e^{-S_f(\theta_f)}\int  dr
\left\langle       \cos(\sqrt{4\pi}\theta(r))\right\rangle,\\       &=
\frac{\Delta}{2\pi}\int   dr\left\lbrace   e^{i\sqrt{4\pi}\theta_s(r)}
\int\mathcal{D}\theta_f                                      e^{\left(
  \int_f\frac{d\omega}{2\pi}[i\sqrt{4\pi}e^{i\omega
      r}\theta_f-|\omega|     \frac{K|\theta_f|^2}{2}]\right)    }+H.c
\right\rbrace       ,\\       &=       \frac{\Delta}{\pi}\int       dr
\cos[\sqrt{4\pi}\theta_s(r)]     e^{\left(     -\frac{1}{K}     \int_f
  \frac{d\omega}{|\omega|} \right) }.\\
\end{aligned}
\end{equation}
We write,  $\int_f \frac{d\omega}{|\omega|}=\int_{\Lambda/b}^{\Lambda}
\frac{d\omega}{|\omega|}=
\ln\Lambda-\ln(\Lambda/b)=\ln[\frac{\Lambda}{\Lambda/b}]=\ln b$.\\

\begin{equation}
\begin{aligned}
\left\langle        S_\Delta(\theta_s,\theta_f)\right\rangle        &=
\frac{\Delta}{\pi}\int   dr   \cos[\sqrt{4\pi}\theta_s(r)]   e^{\left(
  -\frac{1}{K}    \ln    b    \right)   },\\    &=    b^{-\frac{1}{K}}
S_{\Delta}(\theta_s).
\end{aligned}
\end{equation}
Thus the effective action upto  first order cummulant expansion can be
written as,
$$ S_{eff}(\theta_s)= S_s(\theta_s) + b^{-\frac{1}{K}} S_\Delta(\theta_s),$$
\begin{equation}
S_{eff}(\theta_s)=                       \int_{-\Lambda/b}^{\Lambda/b}
\frac{d\omega}{2\pi}|\omega|     \frac{K|\theta_s(\omega)|^2}{2}     +
b^{-\frac{1}{K}}    \int     dr    \left(    \frac{\Delta}{\pi}\right)
\cos[\sqrt{4\pi}\theta_s(r)].
\end{equation}
Now  we  rescale  the  parameters cut-off  momentum  to  the  original
momentum   by  considering,   $\bar{\Lambda}=\frac{\Lambda}{b}$  ,   $
\bar{\omega}=\omega b$ and $\bar{r}=  \frac{r}{b}$. The fields will be
rescaled  as,  $\bar{\theta}(\bar{\omega})=  \frac{\theta(\omega)}{b}$
and we  choose $\bar{\theta}(\bar{r})=\theta_s(r)$. Thus  the rescaled
effective action is given by,
\begin{equation}
\begin{aligned}
S_{eff}(\theta_s)&=                          \int_{-\Lambda}^{\Lambda}
\frac{d\bar{\omega}}{2\pi                   b}\frac{\bar{|\omega|}}{b}
\frac{b^2K|\bar{\theta}(\bar{\omega})|^2}{2} + b^{-\frac{1}{K}} \int b
d\bar{r}                \left(               \frac{\Delta}{\pi}\right)
\cos[\sqrt{4\pi}\bar{\theta}(\bar{r})].
\end{aligned}
\end{equation}
Since we are working in (1+1) dimensional system we have $d^2r= b^2d^2\bar{r}$.
\begin{equation}
\begin{aligned}
S_{eff}(\theta_s)&=                          \int_{-\Lambda}^{\Lambda}
\frac{d\bar{\omega}}{2\pi                   b}\frac{\bar{|\omega|}}{b}
\frac{b^2K|\bar{\theta}(\bar{\omega})|^2}{2}  + b^{-\frac{1}{K}}  \int
b^2         d^2\bar{r}        \left(         \frac{\Delta}{\pi}\right)
\cos[\sqrt{4\pi}\bar{\theta}(\bar{r})],\\ &= \int_{-\Lambda}^{\Lambda}
\frac{d\bar{\omega}}{2\pi               }               \bar{|\omega|}
\frac{K|\bar{\theta}(\bar{\omega})|^2}{2}  +   b^{2-\frac{1}{K}}  \int
d^2\bar{r}               \left(              \frac{\Delta}{\pi}\right)
\cos[\sqrt{4\pi}\bar{\theta}(\bar{r})].
\end{aligned}
\end{equation}
Comparing the coupling constants of rescaled effective action with the
unrenormalized action one can  observe that, $\Delta\rightarrow \Delta
b^{2-\frac{1}{K}}$. Thus we write the RG flow equation as,
$$ \bar{\Delta}= \Delta b^{2-\frac{1}{K}}.$$
We write this equation in the differential form by setting $b=e^{dl}$,

$$\bar{\Delta}=\Delta e^{(2-\frac{1}{K})dl} $$
$$\bar{\Delta}=\Delta[1+(2-\frac{1}{K})dl]     .$$    Defining     the
differential of a parameter as $d\Delta=\bar{\Delta}-\Delta$ we have,
\begin{equation}
\dfrac{d\Delta}{dl}= (2-\frac{1}{K})\Delta.
\end{equation}      
Now we solve for the second order cumulant expansion,
\begin{dmath}
	-\frac{1}{2}(\left\langle
	- S_{\Delta}^2\right\rangle-\left\langle
	- S_{\Delta}\right\rangle^2)=-\frac{{\Delta}^2}{2\pi^2}  \int
	- dr          dr^{\prime}         \left[          \left\langle
	- \cos[\sqrt{4\pi}\theta(r)]
	- \cos[\sqrt{4\pi}\theta(r^{\prime})]            \right\rangle
	- \left\langle     \cos[\sqrt{4\pi}\theta(r)]    \right\rangle
	- \left\langle             \cos[\sqrt{4\pi}\theta(r^{\prime})]
	- \right\rangle \right].
\end{dmath}
First we calculate $\left\langle S_{\Delta}^2\right\rangle $ term. 
\begin{dmath*}
	\left\langle
        S_{\Delta}^2\right\rangle=\frac{{\Delta}^2}{\pi^2}\int      dr
        dr^{\prime}       \left\langle      \cos[\sqrt{4\pi}\theta(r)]
        \cos[\sqrt{4\pi}\theta(r^{\prime})] \right\rangle ,
\end{dmath*}

\begin{dmath*}
	\left\langle
        S_{\Delta}^2\right\rangle=\frac{{\Delta}^2}{4\pi^2}\int     dr
        dr^{\prime}         \left[         \left\langle         \left(
          e^{i\sqrt{4\pi}\theta_s(r)}   e^{i\sqrt{4\pi}\theta_f(r)}  +
          H.c   \right)  \left(   e^{i\sqrt{4\pi}\theta_s(r^{\prime})}
          e^{i\sqrt{4\pi}\theta_f(r^{\prime})}    +     H.c    \right)
          \right\rangle \right],
\end{dmath*}



\begin{dmath}
	\left\langle            S_{\Delta}^2\right\rangle            =
        \frac{{\Delta}^2}{2\pi^2}\int     dr    dr^{\prime}     \left[
          \cos\sqrt{4\pi}[\theta_s(r)+\theta_s(r^{\prime})]     \left(
          e^{-2\pi\left[  \left\langle   \theta_f^2(r)  \right\rangle+
              \left\langle   \theta_f^2(r^{\prime})  \right\rangle   +
              2\left\langle            \theta_f(r)\theta_f(r^{\prime})
              \right\rangle\right]             }\right)\\            +
          \cos\sqrt{4\pi}[\theta_s(r)-\theta_s(r^{\prime})]     \left(
          e^{-2\pi  \left[  \left\langle \theta_f^2(r)  \right\rangle+
              \left\langle   \theta_f^2(r^{\prime})  \right\rangle   -
              2\left\langle            \theta_f(r)\theta_f(r^{\prime})
              \right\rangle\right] }\right)+H.c \right].
\end{dmath}
Now we calculate $\left\langle S_{\Delta}\right\rangle^2$,

\begin{dmath*}
	\left\langle
        S_{\Delta}\right\rangle^2=\frac{{\Delta}^2}{\pi^2}\int      dr
        dr^{\prime}       \left\langle      \cos[\sqrt{4\pi}\theta(r)]
        \right\rangle \left\langle \cos[\sqrt{4\pi}\theta(r^{\prime})]
        \right\rangle ,
\end{dmath*}

\begin{dmath*}
	\left\langle
        S_{\Delta}\right\rangle^2=\frac{{\Delta}^2}{4\pi^2}\int     dr
        dr^{\prime}         \left[         \left\langle         \left(
          e^{i\sqrt{4\pi}\theta_s(r)}   e^{i\sqrt{4\pi}\theta_f(r)}  +
          H.c      \right)\right\rangle       \left\langle      \left(
          e^{i\sqrt{4\pi}\theta_s(r^{\prime})}
          e^{i\sqrt{4\pi}\theta_f(r^{\prime})}    +     H.c    \right)
          \right\rangle \right],
\end{dmath*}


\begin{dmath}
	\left\langle            S_{\Delta}\right\rangle^2            =
        \frac{{\Delta}^2}{2\pi^2}\int     dr    dr^{\prime}     \left[
          \cos\sqrt{4\pi}[\theta_s(r)+\theta_s(r^{\prime})]
          \\    e^{-2\pi\left\langle   \theta^2_f(r)\right\rangle    }
          e^{-2\pi\left\langle \theta^2_f(r^{\prime})\right\rangle } +
          \cos\sqrt{4\pi}[\theta_s(r)-\theta_s(r^{\prime})]
          e^{-2\pi\left\langle       \theta^2_f(r)\right\rangle      }
          e^{-2\pi\left\langle  \theta^2_f(r^{\prime})\right\rangle }+
          H.c \right] .
\end{dmath}
Thus  the  term $(\left\langle  S_{\Delta}^2\right\rangle-\left\langle
S_{\Delta}\right\rangle^2)$ is,
\begin{dmath}
	(\left\langle           S_{\Delta}^2\right\rangle-\left\langle
  S_{\Delta}\right\rangle^2)=     \frac{{\Delta}^2}{2\pi^2}\int     dr
  dr^{\prime} \left[ \cos\sqrt{4\pi}[\theta_s(r)+\theta_s(r^{\prime})]
    \left(  e^{-2\pi\left[  \left\langle \theta_f^2(r)  \right\rangle+
        \left\langle     \theta_f^2(r^{\prime})    \right\rangle     +
        2\left\langle                  \theta_f(r)\theta_f(r^{\prime})
        \right\rangle\right]                 }\right)                +
    \cos\sqrt{4\pi}[\theta_s(r)-\theta_s(r^{\prime})]  \left( e^{-2\pi
      \left[  \left\langle  \theta_f^2(r) \right\rangle+  \left\langle
        \theta_f^2(r^{\prime})    \right\rangle     -    2\left\langle
        \theta_f(r)\theta_f(r^{\prime}) \right\rangle\right] }\right)+
    H.c \right] -  \frac{{\Delta}^2}{2\pi^2}\int dr dr^{\prime} \left[
    \cos\sqrt{4\pi}[\theta_s(r)+\theta_s(r^{\prime})]
    e^{-2\pi\left\langle          \theta^2_f(r)\right\rangle         }
    e^{-2\pi\left\langle   \theta^2_f(r^{\prime})\right\rangle   }   +
    \cos\sqrt{4\pi}[\theta_s(r)-\theta_s(r^{\prime})]
    e^{-2\pi\left\langle          \theta^2_f(r)\right\rangle         }
    e^{-2\pi\left\langle  \theta^2_f(r^{\prime})\right\rangle  }+  H.c
    \right] ,
\end{dmath}

\begin{dmath}
	(\left\langle           S_{\Delta}^2\right\rangle-\left\langle
  S_{\Delta}\right\rangle^2)=     \frac{{\Delta}^2}{2\pi^2}\int     dr
  dr^{\prime} \left[ \cos\sqrt{4\pi}[\theta_s(r)+\theta_s(r^{\prime})]
    \left(  e^{-2\pi\left[  \left\langle \theta_f^2(r)  \right\rangle+
        \left\langle     \theta_f^2(r^{\prime})    \right\rangle     +
        2\left\langle                  \theta_f(r)\theta_f(r^{\prime})
        \right\rangle\right]       }      -       e^{-2\pi\left\langle
      \theta^2_f(r)\right\rangle         }        e^{-2\pi\left\langle
      \theta^2_f(r^{\prime})\right\rangle         }\right)         \\+
    \cos\sqrt{4\pi}[\theta_s(r)+\theta_s(r^{\prime})]  \left( e^{-2\pi
      \left[  \left\langle  \theta_f^2(r) \right\rangle+  \left\langle
        \theta_f^2(r^{\prime})    \right\rangle     -    2\left\langle
        \theta_f(r)\theta_f(r^{\prime})           \right\rangle\right]
    }-e^{-2\pi\left\langle         \theta^2_f(r)\right\rangle        }
    e^{-2\pi\left\langle  \theta^2_f(r^{\prime})\right\rangle }\right)
    \right] .
\end{dmath}
Thus,
\begin{dmath}
-\frac{1}{2}(\left\langle       S_{\Delta}^2\right\rangle-\left\langle
S_{\Delta}\right\rangle^2)   =   -\frac{{\Delta}^2}{4\pi^2}  \int   dr
dr^{\prime}                          \\                         \left[
  \cos\sqrt{4\pi}[\theta_s(r)+\theta_s(r^{\prime})]
  e^{-2\pi[\left\langle  \theta_f^2(r)  \right\rangle  +  \left\langle
      \theta_f^2(r^{\prime})           \right\rangle]}          \left(
  e^{-4\pi\left\langle \theta_f(r)\theta_f(r^{\prime}) \right\rangle }
  -1   \right)   +   \cos\sqrt{4\pi}[\theta_s(r)-\theta_s(r^{\prime})]
  e^{-2\pi[\left\langle  \theta_f^2(r)  \right\rangle  +  \left\langle
      \theta_f^2(r^{\prime})           \right\rangle]}          \left(
  e^{4\pi\left\langle \theta_f(r)\theta_f(r^{\prime})  \right\rangle }
  -1 \right) \right] . \label{seconddelta}
\end{dmath}
The correlation function $\left\langle \theta_f(r)\theta_f(r^{\prime})
\right\rangle$ is calculated as,
\begin{equation}
\left\langle         \theta_f(r)\theta_f(r^{\prime})\right\rangle=\int
\mathcal{D}\theta_f  e^{\left[  -\int_f\frac{d\omega}{2\pi}  K|\omega|
    |\theta_f|^2\right] } \theta_f(r) \theta_f(r^{\prime}),
\end{equation}
\begin{equation}
\left\langle         \theta_f(r)\theta_f(r^{\prime})\right\rangle=\int
\mathcal{D}\theta_f  e^{\left[  -\int_f\frac{d\omega}{2\pi}  K|\omega|
    |\theta_f|^2\right]   }   \int_f\frac{d\omega}{2\pi}   e^{-i\omega
  r}\theta(\omega)                 \int_f\frac{d\omega^{\prime}}{2\pi}
e^{-i\omega^{\prime}r^{\prime}}\theta(\omega^{\prime}) ,
\end{equation}
\begin{equation}
\left\langle\theta_f(r)\theta_f(r^{\prime})\right\rangle=\int_f\frac{d\omega
  d\omega^{\prime}}{(2\pi)^2}                             e^{-i(\omega
  r+\omega^{\prime}r^{\prime})}  e^{\left[ -\int_f\frac{d\omega}{2\pi}
    K|\omega||\theta_f|^2\right]            }           \theta(\omega)
\theta(\omega^{\prime}),
\end{equation}
\begin{equation}
\left\langle       \theta_f(r)\theta_f(r^{\prime})\right\rangle\propto
\int_f\frac{d\omega        d\omega^{\prime}}{2\pi}        e^{-i(\omega
  r+\omega^{\prime}r^{\prime})}
\frac{1}{2|\omega|K}\delta(\omega+\omega^{\prime}),
\end{equation}
\begin{equation}
\left\langle \theta_f(r)\theta_f(r^{\prime})\right\rangle=\frac{1}{2K}
\int_{\Lambda/b<|\omega|<\Lambda}
\frac{d|\omega|}{2\pi}e^{-i\omega(r-r^{\prime})}|\omega|^{-1},
\end{equation}
\begin{equation}
\left\langle
\theta_f(r)\theta_f(r^{\prime})\right\rangle=\frac{1}{2\pi
  K}\int_{\Lambda/b}^{\Lambda}\frac{d\omega}{\omega}e^{-i\omega(r-r^{\prime})}.
\end{equation}
For $r^{\prime}\rightarrow r$ we will have,
\begin{equation}
\left\langle  \theta_f(r)\theta_f(r^{\prime})   \right\rangle  \approx
\left\langle      \theta_f^2(r)\right\rangle      =      \frac{1}{2\pi
  K}\int_{\Lambda/b}^{\Lambda}  \frac{d\omega}{\omega}=  \frac{1}{2\pi
  K} \ln b .
\end{equation}
We introduce  the relative  coordinate $s=r-r^{\prime}$ and  center of
mass coordinate $T=(r+r^{\prime})/2$. Thus we have,
\begin{equation*}
\cos\sqrt{4\pi}[\theta_s(r)+\theta_s(r^{\prime})]=\cos [4\sqrt{\pi}(\theta_s(T))].
\end{equation*} 
This term is RG irrelevent term. For small $s$ cosine can be approximated by, 
\begin{equation*}
\cos\sqrt{4\pi}[\theta_s(r)-\theta_s(r^{\prime})]=     1    -     2\pi
(s\partial_T\theta_s(T))^2.
\end{equation*}
Thus eq.\ref{seconddelta} can be written as,
\begin{equation}
\begin{aligned}
  -\frac{1}{2}(\left\langle     S_{\Delta}^2\right\rangle-\left\langle
  S_{\Delta}\right\rangle^2) 
&=-\frac{{\Delta}^2}{4\pi^2}      \left(     1-      \left(\frac{1}{b}
\right)^{\frac{2}{K}} \right)  \int_{0}^{b/\Lambda} ds  \int dT (  1 -
2\pi (s\partial_T\theta_s(T))^2).
\end{aligned}
\end{equation}
Here the  first term turns out  to be field independent  term. Thus we
consider only second term,
\begin{equation}
\begin{aligned}
-\frac{1}{2}(\left\langle       S_{\Delta}^2\right\rangle-\left\langle
S_{\Delta}\right\rangle^2)&=\frac{{\Delta}^2}{4\pi^2}     \left(    1-
\left(\frac{1}{b}  \right)^{\frac{2}{K}} \right)  \int_{0}^{b/\Lambda}
ds                   \int                   dT                   (2\pi
(s\partial_T\theta_s(T))^2),\\  &=\frac{{\Delta}^2}{2\pi}   \left(  1-
\left(\frac{1}{b}  \right)^{\frac{2}{K}} \right)  \int_{0}^{b/\Lambda}
s^2 ds \int dT (\partial_T\theta_s(T))^2,\\
&=\frac{{\Delta}^2}{6\pi\Lambda^3}       \left(      \left(\frac{1}{b}
\right)^{-3}-   \left(\frac{1}{b}    \right)^{\frac{2}{K}-3}   \right)
\int_{-\Lambda/b}^{\Lambda/b}             \frac{d\omega}{2\pi}|\omega|
\frac{K|\theta_s(\omega)|^2}{2}.
\end{aligned}
\end{equation}
After rescaling the parameters and fields the equation will have the form,
\begin{equation}
-\frac{1}{2}(\left\langle       S_{\Delta}^2\right\rangle-\left\langle
S_{\Delta}\right\rangle^2)=\frac{{\Delta}^2}{6\pi\Lambda^3}     \left(
\left(\frac{1}{b}            \right)^{-3}-           \left(\frac{1}{b}
\right)^{\frac{2}{K}-3}        \right)       \int_{-\Lambda}^{\Lambda}
\frac{d\bar{\omega}}{2\pi               }               \bar{|\omega|}
\frac{K|\bar{\theta}(\bar{\omega})|^2}{2}.
\end{equation}
Now the effective action can be written as, 
\begin{equation}
\begin{split}
	S_{eff}= \int_{-\Lambda}^{\Lambda} \frac{d\bar{\omega}}{2\pi }
        \bar{|\omega|}   \frac{K|\bar{\theta}(\bar{\omega})|^2}{2}   +
        b^{2-\frac{1}{K}}        \int         d^2\bar{r}        \left(
        \frac{{\Delta}}{\pi}\right)
        \cos[\sqrt{4\pi}\bar{\theta}(\bar{r})]\\ +\frac{{\Delta}^2}{6\pi\Lambda
          ^3} \left( \left(\frac{1}{b} \right)^{-3}- \left(\frac{1}{b}
        \right)^{\frac{2}{K}-3}    \right)   \int_{-\Lambda}^{\Lambda}
        \frac{d\bar{\omega}}{2\pi           }           \bar{|\omega|}
        \frac{K|\bar{\theta}(\bar{\omega})|^2}{2},
\end{split}
\end{equation}

\begin{equation}
S_{eff}    =    \left[    1+\frac{{\Delta}^2}{6\pi\Lambda^3}    \left(
  \left(\frac{1}{b}           \right)^{-3}-          \left(\frac{1}{b}
  \right)^{\frac{2}{K}-3}      \right)     \right]      S_s(\theta_s)+
b^{2-\frac{1}{K}}S_{\Delta}(\theta_s).
\end{equation}      
Comparing the  rescaled effective action  with the original  action we
obtain RG flow equation,
\begin{equation}
{\bar{K}}=    K\left[     1+\frac{{\Delta}^2}{6\pi\Lambda^3}    \left(
  \left(\frac{1}{b}           \right)^{-3}-          \left(\frac{1}{b}
  \right)^{\frac{2}{K}-3} \right) \right] .
\end{equation}
 Defining the  differential of  a parameter  as $dK=\bar{K}-K$  and by
 setting $b=e^{dl}$ we get RG flow equation in the differential form,
 $$ dK= -\frac{K\Delta^2}{6\pi\Lambda^3}(e^{3dl}-e^{(3-\frac{2}{K})dl}),$$
 $$ dK= -\frac{K\Delta^2}{6\pi\Lambda^3} [1+3dl-1-3dl+(\frac{2}{K})dl],$$
 \begin{equation}
 \dfrac{dK}{dl}=\left(\frac{{\Delta}^2}{3\pi\Lambda^3}\right).
 \end{equation}
Now we rescale ${\Delta} \rightarrow {\Delta}\sqrt{\frac{1}{3\pi\Lambda^3}}$.
Thus RG flow equations of $H_1$ is given by equation,
\begin{equation}
\dfrac{d{\Delta}}{dl}= \left( 2-\frac{1}{K}\right) {\Delta}, \;\;\;\;\;\;\;
\frac{d K}{dl} =  {\Delta}^2 .
\end{equation}         
One can follow  the same procedure to obtain the  RG equations for the
Hamiltonian $H_2$.\\

\textbf{\large B) Derivation of Quantum BKT equations for finite $\mu$}\\
We start with the Bosonized model Hamiltonian $H$ with $g_u=0$,
	\begin{equation}
	\begin{split}
	H  =  \frac{v}{2} \int \left[ \frac{1}{K} ( {({\partial_x \phi(x) })}^2 + 
	K {({\partial_x  \theta(x) })}^2  )\right] dx - \left( \frac{\mu}{\sqrt{\pi}}\right) \int \partial_x \phi(x) dx 
	+ \frac{B}{\pi} \int cos(\sqrt{4 \pi} \phi(x)) dx\\ - \frac{\Delta}{\pi} \int cos(\sqrt{4 \pi \theta(x)}) dx.
	\end{split}
	\end{equation}
	After rescaling the fields as, $\phi^{\prime} = \frac{\phi}{\sqrt{K}}$ and $ \theta^{\prime} = \sqrt{K}\theta$, one can write,
\begin{equation}
	\begin{split}
	H  =  \frac{v}{2} \int \left[ {({\partial_x \phi^{\prime} })}^2 + 
	{({\partial_x  \theta^{\prime} })}^2  \right] dx - \mu \sqrt{\frac{K}{\pi}} \int \partial_x \phi^{\prime} dx+ \frac{B}{\pi} \int cos(\sqrt{4 \pi K} \phi^{\prime}) dx\\ - \frac{\Delta}{\pi} \int cos(\sqrt{\frac{4 \pi}{K} \theta^{\prime}}) dx.
	\end{split}
	\end{equation}
Writing the Lagrangian using the Hmailton's equations, $ \partial_{x} \theta^{\prime} = -\frac{1}{v} \partial_{t} \phi^{\prime}$ and $ \partial_{x}\phi^{\prime} = -\frac{1}{v} \partial_{t} \theta^{\prime}$ leads us to,
\begin{equation}
\begin{aligned}
\mathcal{L}_0&= \Pi_{\phi^{\prime}} \partial_t \phi^{\prime}-H_0^{\prime},\\
&= \left( \frac{1}{v}\partial_t \phi^{\prime} \right) \partial_t \phi^{\prime} -\frac{v}{2}(\partial_x \theta^{\prime})^2- \frac{v}{2}(\partial_x \phi^{\prime})^2,\\
&= \frac{1}{2}[v^{-1}(\partial_t \phi^{\prime})^2-v(\partial_x \phi^{\prime})^2].
\end{aligned}
\end{equation}
Here the Lagrangian $\mathcal{L}_0$, is written in terms of $\phi^{\prime}$ field. One can also write the $\mathcal{L}_0$ in terms of $\theta^{\prime}$
field as,
\begin{equation}
\mathcal{L}_0 = \frac{1}{2}[v^{-1}(\partial_t \theta^{\prime})^2-v(\partial_x \theta^{\prime})^2].
\end{equation}
Putting these two together one can have $\mathcal{L}_0$ in terms of both $\phi^{\prime}$ and $\theta^{\prime}$,
\begin{equation}
\mathcal{L}_0 = \frac{1}{4} [v^{-1}(\partial_t \phi^{\prime})^2-v(\partial_x \phi^{\prime})^2 + v^{-1}(\partial_t \theta^{\prime})^2-v(\partial_x \theta^{\prime})^2].
\end{equation}
Writting the $\mathcal{L}_0$ in imaginary time i.e, $\tau=it$,
\begin{equation}
\mathcal{L}_0 = \frac{1}{4} [v^{-1}(i\partial_{\tau} \phi^{\prime})^2 + v^{-1}(i\partial_{\tau} \theta^{\prime})^2-v(\partial_x \phi^{\prime})^2 -v(\partial_x \theta^{\prime})^2],
\end{equation}
\begin{equation}
\mathcal{L}_0 = -\frac{1}{4} [v^{-1}(\partial_{\tau} \phi^{\prime})^2 + v^{-1}(\partial_{\tau} \theta^{\prime})^2 + v(\partial_x \phi^{\prime})^2 + v(\partial_x \theta^{\prime})^2].
\end{equation}
Lagrangian of the interaction ($\mathcal{L}_{int}$) terms can be obtained by the relation $\mathcal{L}_{int}=-H_{int}$. 
\begin{equation}
\begin{aligned}
\mathcal{L}_{int} 
&= \mu \sqrt{\frac{K}{\pi}}  \partial_x \phi^{\prime} - \frac{B}{\pi} cos(\sqrt{4 \pi K} \phi^{\prime}) + \frac{\Delta}{\pi} cos(\sqrt{\frac{4 \pi}{K}} \theta^{\prime}), \\
&= -\frac{\mu}{v} \sqrt{\frac{K}{\pi}}  \partial_t \theta^{\prime} - \frac{B}{\pi} cos(\sqrt{4 \pi K} \phi^{\prime}) + \frac{\Delta}{\pi} cos(\sqrt{\frac{4 \pi}{K}} \theta^{\prime}). 
\end{aligned}
\end{equation}
Writting this   in imaginary time $\tau=it$,
\begin{equation}
\mathcal{L}_{int} = -\frac{i\mu}{v} \sqrt{\frac{K}{\pi}}  \partial_{\tau} \theta^{\prime} - \frac{B}{\pi} cos(\sqrt{4 \pi K} \phi^{\prime}) + \frac{\Delta}{\pi} cos(\sqrt{\frac{4 \pi}{K}} \theta^{\prime}). 
\end{equation}
The Euclidean action can be written as, $S_E= -\int dr \mathcal{L} = -\int dr (\mathcal{L}_0 + \mathcal{L}_{int})$, where $r=(\tau,x)$.
Thus the partition function can be derived as
\begin{dmath}
\mathcal{Z}= \int \mathcal{D}\phi \mathcal{D}\theta exp\left[ - \int_{-\Lambda}^{\Lambda} \frac{d\omega}{2\pi}|\omega| \left( \frac{|\phi(\omega)|^2}{2K} + \frac{K|\theta (\omega)|^2}{2}\right) \\
- \int dr \left( \frac{i\mu}{v\sqrt{\pi}}  (\partial_{r} \theta) + \frac{B}{\pi} cos(\sqrt{4 \pi } \phi) - \frac{\Delta}{\pi} cos(\sqrt{4 \pi} \theta)\right)  \right]. 
\end{dmath}
Now we divide the fields into slow and fast modes and integrate out the fast modes. The filed $\phi$ is $\phi(r)=\phi_s(r)+\phi_f(r)$ similarly the field $\theta$ is $\theta(r)=\theta_s(r)+\theta_f(r)$ where,
\begin{equation}
\phi_s(r)=\int_{-\Lambda/b}^{\Lambda/b} \frac{d\omega}{2\pi} e^{-i\omega r}\phi(\omega) \;\;\;\;\&\;\;\; \phi_f(r)=\int_{\Lambda/b<|\omega_n|<\Lambda} \frac{d\omega}{2\pi} e^{-i\omega r}\phi(\omega),
\end{equation}
\begin{equation}
\theta_s(r)=\int_{-\Lambda/b}^{\Lambda/b} \frac{d\omega}{2\pi} e^{-i\omega r}\theta(\omega) \;\;\;\;\&\;\;\; \theta_f(r)=\int_{\Lambda/b<|\omega_n|<\Lambda} \frac{d\omega}{2\pi} e^{-i\omega r}\theta(\omega).
\end{equation}
Thus the partition function $\mathcal{Z}$ is,
\begin{equation}
\mathcal{Z}=\int \mathcal{D}\phi_s \mathcal{D}\phi_f \mathcal{D}\theta_s \mathcal{D}\theta_f e^{-S_s(\phi_s,\theta_s)} e^{-S_f(\phi_f,\theta_f)} e^{-S_{int}(\phi,\theta)}.
\end{equation}
Using the relation $\left\langle A \right\rangle_f = \int \mathcal{D}\phi_f e^{-S_f(\phi_f,\theta_f)} A $, one can write,
\begin{equation}
\mathcal{Z}=\int \mathcal{D}\phi_s \mathcal{D}\theta_s e^{-S_s(\phi_s,\theta_s)} \left\langle e^{-S_{int}(\phi,\theta)} \right\rangle_f.
\end{equation}
We write the effective action as,
\begin{equation}
e^{-S_{eff}(\phi_s,\theta_s)}=e^{-S_s(\phi_s,\theta_s)} \left\langle e^{-S_{int}(\phi,\theta)} \right\rangle_f.
\end{equation}
Taking $\ln$ on both side gives,
\begin{equation}
S_{eff}(\phi_s,\theta_s)= S_s(\phi_s,\theta_s) - \ln\left\langle e^{-S_{int}(\phi,\theta)} \right\rangle_f.
\end{equation}
By writing the cumulant expansion up to second order, we have
\begin{equation}
S_{eff}(\phi_s,\theta_s)= S_s(\phi_s,\theta_s)+ \left\langle S_{int}(\phi,\theta) \right\rangle_f - \frac{1}{2} \left( \left\langle S^2_{int}(\phi,\theta) \right\rangle_f - \left\langle S_{int}(\phi,\theta) \right\rangle^2_f \right) .
\end{equation}
Now we calculate the first order cumulant expansion.
\begin{equation}
\begin{split} 
\left\langle S_{int}(\phi,\theta) \right\rangle_f = \int dr \left\langle\frac{i\mu}{v\sqrt{\pi}} \partial_{r} \theta(r) \right\rangle_f + \int dr \frac{B}{\pi}\left\langle  \cos(\sqrt{4\pi}\phi(r))\right\rangle_f -\int dr \frac{\Delta}{\pi}\left\langle  \cos(\sqrt{4\pi}\theta(r))\right\rangle_f .\\ 
\end{split}
\end{equation}
The second term is,
\begin{equation}
\begin{aligned}
\int dr \frac{B}{\pi}\left\langle  \cos(\sqrt{4\pi}\phi(r))\right\rangle_f &= \int dr \left( \frac{B}{\pi}\right) \int \mathcal{D}\phi_f e^{-S_f[\phi_f]}   \cos(\sqrt{4\pi}\phi(r)), \\
&=\frac{B}{2\pi} \int dr \left\lbrace e^{i\sqrt{4\pi}\phi_s(r)} \int \mathcal{D}\phi_f e^{\int_f \frac{d\omega}{2\pi} \left[i\sqrt{4\pi} e^{i\omega r}\phi_f(\omega) - |\omega|  \frac{|\phi_f(\omega)|^2}{2K} \right]}  + H.c \right\rbrace, \\
&=\frac{B}{\pi} \int dr \cos[\sqrt{4\pi}\phi_s(r)] e^{-K\int_f\frac{d\omega}{|\omega|}},\\
&= \frac{B}{\pi} \int dr \cos[\sqrt{4\pi}\phi_s(r)] e^{\ln b^{-K}},\\
&= \frac{B}{\pi} (b^{-K}) \int dr \cos[\sqrt{4\pi}\phi_s(r)] .
\end{aligned}
\end{equation}
Thus we have,
\begin{equation}
\int dr \frac{B}{\pi}\left\langle  \cos(\sqrt{4\pi}\phi(r))\right\rangle_f =  b^{-K}  \int dr \left( \frac{B}{\pi} \right) \cos[\sqrt{4\pi}\phi_s(r)].
\end{equation}
Following the above porcedure one can arrive at the following equations for $\Delta$, 
\begin{equation}
\int dr \frac{\Delta}{\pi}\left\langle  \cos(\sqrt{4\pi}\theta(r))\right\rangle_f = b^{-\frac{1}{K}}\int dr \left( \frac{\Delta}{\pi} \right) \cos[\sqrt{4\pi}\theta_s(r)].
\end{equation}
The first order cumulant expansion can be re-written as,
\begin{equation}
\begin{split}
\left\langle S_{int}(\phi,\theta) \right\rangle_f=b^{-K}\int dr \left( \frac{B}{\pi} \cos(\sqrt{4\pi}\phi_s(r))\right) - b^{-\frac{1}{K}}\int dr \left( \frac{\Delta}{\pi}\cos(\sqrt{4\pi}\theta_s(r))\right). \\ 
\end{split}
\end{equation}
Now we calculate the second order cumulant expansion which has the following terms,
\begin{dmath}
-\frac{1}{2}(\left\langle S_{int}^2\right\rangle-\left\langle S_{int}\right\rangle^2) 
= -\frac{1}{2} \int dr dr^{\prime}  \left( -\frac{\mu^2}{v^2\pi} \right) \left(\left\langle \partial_{r}\phi(r)\partial_{r^{\prime}}\phi(r^{\prime})\right\rangle 
- \left\langle \partial_{r}\phi(r) \right\rangle \left\langle \partial_{r^{\prime}}\phi(r^{\prime})\right\rangle\right) 
-\frac{1}{2} \int dr dr^{\prime}\left( \frac{B^2}{\pi^2}\right) \left(  \left\langle \cos(\sqrt{4\pi}\phi(r)) \cos(\sqrt{4\pi}\phi(r^{\prime}))\right\rangle 
- \left\langle \cos(\sqrt{4\pi}\phi(r)) \right\rangle \left\langle \cos(\sqrt{4\pi}\phi(r^{\prime}))\right\rangle \right)
 -\frac{1}{2} \int dr dr^{\prime} \left( \frac{\Delta^2}{\pi^2}\right) \left(  \left\langle \cos(\sqrt{4\pi}\theta(r)) \cos(\sqrt{4\pi}\theta(r^{\prime}))\right\rangle 
- \left\langle \cos(\sqrt{4\pi}\theta(r)) \right\rangle \left\langle \cos(\sqrt{4\pi}\theta(r^{\prime}))\right\rangle \right)  
-\frac{1}{2} \int dr dr^{\prime}\left( \frac{i\mu B}{v\sqrt{\pi}\pi} \right) \left( \left\langle  \partial_{r}\phi(r)\cos(\sqrt{4\pi}\phi(r^{\prime}))\right\rangle 
\left\langle \partial_{r}\phi(r)\right\rangle - \left\langle \cos(\sqrt{4\pi}\phi(r^{\prime})) \right\rangle \right)\\
-\frac{1}{2} \int dr dr^{\prime}\left(- \frac{ i\mu \Delta}{v\sqrt{\pi}\pi}\right) \left(  \left\langle  \partial_{r}\phi(r) \cos(\sqrt{4\pi}\theta(r^{\prime}))\right\rangle - \left\langle  \partial_{r}\phi(r)\right\rangle \left\langle \cos(\sqrt{4\pi}\theta(r^{\prime}))\right\rangle \right)
-\frac{1}{2} \int dr dr^{\prime}\left( \frac{Bi\mu}{\pi v\sqrt{\pi}}\right) \left(  \left\langle \cos(\sqrt{4\pi}\phi(r)) \partial_{r^{\prime}}\phi(r^{\prime})\right\rangle 
- \left\langle \cos(\sqrt{4\pi}\phi(r)) \right\rangle \left\langle \partial_{r^{\prime}}\phi(r^{\prime})\right\rangle \right)
 -\frac{1}{2} \int dr dr^{\prime}\left( -\frac{B\Delta}{\pi^2}\right) \left(  \left\langle \cos(\sqrt{4\pi}\phi(r)) \cos(\sqrt{4\pi}\theta(r^{\prime}))\right\rangle 
- \left\langle \cos(\sqrt{4\pi}\phi(r)) \right\rangle \left\langle \cos(\sqrt{4\pi}\theta(r^{\prime}))\right\rangle \right)
-\frac{1}{2} \int dr dr^{\prime}\left(- \frac{\Delta i \mu}{\pi v\sqrt{\pi}}\right) \left(  \left\langle \cos(\sqrt{4\pi}\theta(r)) \partial_{r^{\prime}}\phi(r^{\prime})\right\rangle 
- \left\langle \cos(\sqrt{4\pi}\theta(r)) \right\rangle \left\langle \partial_{r^{\prime}}\phi(r^{\prime})\right\rangle \right)
-\frac{1}{2} \int dr dr^{\prime}\left( -\frac{\Delta B}{\pi^2}\right) \left(  \left\langle \cos(\sqrt{4\pi}\theta(r)) \cos(\sqrt{4\pi}\phi(r^{\prime}))\right\rangle 
- \left\langle \cos(\sqrt{4\pi}\theta(r)) \right\rangle \left\langle \cos(\sqrt{4\pi}\phi(r^{\prime}))\right\rangle \right).
\end{dmath}
The $B^2$ term can be written as, 
\begin{dmath}
-\frac{1}{2} \int dr dr^{\prime}\left( \frac{B^2}{\pi^2} \left\langle \cos(\sqrt{4\pi}\phi(r)) \cos(\sqrt{4\pi}\phi(r^{\prime}))\right\rangle 
- \left\langle \cos(\sqrt{4\pi}\phi(r)) \right\rangle \left\langle \cos(\sqrt{4\pi}\phi(r^{\prime}))\right\rangle \right) = \frac{B^2}{4\pi^2}\left(1- b^{-2K}\right)\int dr (\partial_{r}\phi_s)^2 - \frac{B^2}{2\pi^2}\left(b^{-4K}- b^{-2K}\right)\int dr \cos[\sqrt{16\pi}\phi_s(r)].
\end{dmath}
Similarly one can write,
\begin{dmath}
-\frac{1}{2} \int dr dr^{\prime} \left( \frac{\Delta^2}{\pi^2} \left\langle \cos(\sqrt{4\pi}\theta(r)) \cos(\sqrt{4\pi}\theta(r^{\prime}))\right\rangle 
- \left\langle \cos(\sqrt{4\pi}\theta(r)) \right\rangle \left\langle \cos(\sqrt{4\pi}\theta(r^{\prime}))\right\rangle \right) = \frac{\Delta^2}{4\pi^2}\left(1- b^{-\frac{2}{K}}\right)\int dr (\partial_{r}\theta_s)^2.
\end{dmath}
Now we calculate $\Delta i\mu$ term, 
\begin{dmath}
-\frac{1}{2} \int dr dr^{\prime}\left(- \frac{\Delta i \mu}{\pi v\sqrt{\pi}} \left\langle \cos(\sqrt{4\pi}\theta(r)) \partial_{r^{\prime}}\theta (r^{\prime})\right\rangle 
- \left\langle \cos(\sqrt{4\pi}\theta(r)) \right\rangle \left\langle \partial_{r^{\prime}}\theta (r^{\prime})\right\rangle \right) = \frac{\Delta i \mu}{2\pi v\sqrt{\pi}}\int dr dr^{\prime}\left[ \left\langle \cos[\sqrt{4\pi}\theta(r)](\partial_{r^{\prime}}\theta_s(r^{\prime})+\partial_{r^{\prime}}\theta_f(r^{\prime}))\right\rangle - \left\langle \cos[\sqrt{4\pi}\theta(r)] \right\rangle \left\langle \partial_{r^{\prime}}\theta_s(r^{\prime})+\partial_{r^{\prime}}\theta_f(r^{\prime}) \right\rangle \right], \\
=\frac{\Delta i \mu}{2\pi v\sqrt{\pi}}\int dr dr^{\prime}\left[\left\langle \cos[\sqrt{4\pi}\theta(r)]\partial_{r^{\prime}}\theta_f(r^{\prime})\right\rangle \right] .
\end{dmath} 
The correlation function $\left\langle \cos[\sqrt{4\pi}\theta(r)]\partial_{r^{\prime}}\theta_f(r^{\prime})\right\rangle$ can be written as,
\begin{equation}
\left\langle \cos[\sqrt{4\pi}\theta(r)]\partial_{r^{\prime}}\theta_f(r^{\prime})\right\rangle = -2\sqrt{\pi} \sin[\sqrt{4\pi}\theta_s(r)]\partial_{r^{\prime}}\left\langle \theta_f(r^{\prime})\theta_f(r) \right\rangle e^{-2\pi\left\langle \theta^2_f(r) \right\rangle }.
\end{equation}
Thus we have,
\begin{equation}
\begin{aligned}
&=-\frac{\Delta i \mu}{\pi v}\int dr dr^{\prime} \sin[\sqrt{4\pi}\theta_s(r)]\partial_{r^{\prime}}\left\langle \theta_f(r^{\prime})\theta_f(r) \right\rangle e^{-2\pi\left\langle \theta^2_f(r) \right\rangle },\\
&=\frac{\Delta i \mu}{2\pi^2 v}\int dr  \sin[\sqrt{4\pi}\theta_s(r)](e^{\frac{1}{ K}\ln b}-1)( e^{-\frac{1}{ K}\ln b}),\\
&\approx-\frac{ \Delta \mu}{2\pi^2 v}(1-b^{-\frac{1}{K}})\int dr \cos[\sqrt{4\pi}\theta_s(r)].
\end{aligned}
\end{equation}
Thus combined $\Delta i\mu$ and $i\mu\Delta$ terms gives,
\begin{dmath}
-\frac{1}{2} \int dr dr^{\prime}\left(- \frac{\Delta i \mu}{\pi v\sqrt{\pi}} \left\langle \cos(\sqrt{4\pi}\theta(r)) \partial_{r^{\prime}}\theta (r^{\prime})\right\rangle 
- \left\langle \cos(\sqrt{4\pi}\theta(r)) \right\rangle \left\langle \partial_{r^{\prime}}\theta (r^{\prime})\right\rangle \right)=-\frac{ \Delta \mu}{\pi^2 v}(1-b^{-\frac{1}{K}})\int dr \cos[\sqrt{4\pi}\theta_s(r)].
\end{dmath}
In the case of $Bi\mu$ the correlation function $\left\langle \phi_f(r) \theta_f(r^{\prime}) \right\rangle $ is,
\begin{equation}
\begin{aligned}
\left\langle \phi_f(r) \theta_f(r^{\prime}) \right\rangle &= \left\langle (\phi_{R\uparrow}+\phi_{L\downarrow})(\phi_{R\uparrow}^{\prime}-\phi_{L\downarrow}^{\prime}) \right\rangle, \\
&= \left\langle \phi_{R\uparrow}\phi_{R\uparrow}^{\prime} - \phi_{R\uparrow}\phi_{L\downarrow}^{\prime} + \phi_{L\downarrow} \phi_{R\uparrow}^{\prime} - \phi_{L\downarrow}\phi_{L\downarrow}^{\prime} \right\rangle,\\
&=0 .
\end{aligned}
\end{equation}
Thus the combined $Bi\mu$ and $i\mu B$ terms equals to $0$.
\begin{dmath}
-\frac{1}{2} \int dr dr^{\prime}\left( -\frac{Bi\mu}{\pi v\sqrt{\pi}} \left\langle \cos(\sqrt{4\pi}\phi(r)) \partial_{r^{\prime}}\phi(r^{\prime})\right\rangle 
- \left\langle \cos(\sqrt{4\pi}\phi(r)) \right\rangle \left\langle \partial_{r^{\prime}}\phi(r^{\prime})\right\rangle \right) = 0.
\end{dmath}
Now we calculate the term $B\Delta$, 
\begin{dmath}
-\frac{1}{2} \int dr dr^{\prime}\left( -\frac{B\Delta}{\pi^2} \left\langle \cos(\sqrt{4\pi}\phi(r)) \cos(\sqrt{4\pi}\theta(r^{\prime}))\right\rangle 
- \left\langle \cos(\sqrt{4\pi}\phi(r)) \right\rangle \left\langle \cos(\sqrt{4\pi}\theta(r^{\prime}))\right\rangle \right) = \frac{B\Delta}{4\pi^2} \int dr dr^{\prime}  \left[ \cos\sqrt{4\pi}[\phi_s(r)+\theta_s(r^{\prime})]  \left( e^{-2\pi\left\langle \left( \phi_f(r)  +  \theta_f(r^{\prime})\right)^2 \right\rangle} - e^{-2\pi[\left\langle \phi_f^2(r) \right\rangle + \left\langle \theta_f^2(r^{\prime}) \right\rangle]}   \right) + \cos\sqrt{4\pi}[\phi_s(r)-\theta_s(r^{\prime})] \left( e^{-2\pi\left\langle \left( \phi_f(r)  -  \theta_f(r^{\prime})\right)^2 \right\rangle} - e^{-2\pi[\left\langle \phi_f^2(r) \right\rangle + \left\langle \theta_f^2(r^{\prime}) \right\rangle]} \right) \right] .
\end{dmath}
Here the correlation function is,
\begin{equation*}
e^{-2\pi\left\langle \left( \phi_f(r)  \pm  \theta_f(r^{\prime})\right)^2 \right\rangle} = e^{-2\pi \left\langle \phi_f^2(r) \right\rangle + \left\langle \theta_f^2(r^{\prime})\right\rangle  \pm 2\left\langle \phi_f(r)\theta_f(r^{\prime})\right\rangle }.
\end{equation*}
We know that $\left\langle \phi_f(r)\theta_f(r^{\prime})\right\rangle =0$. Thus we have, 
\begin{equation}
e^{-2\pi\left\langle \left( \phi_f(r)  \pm  \theta_f(r^{\prime})\right)^2 \right\rangle} = e^{-2\pi[\left\langle \phi_f^2(r) \right\rangle + \left\langle \theta_f^2(r^{\prime}) \right\rangle]}.
\end{equation}
These two exponentials cancel each other making the whole term $0$. Thus the combined $B\Delta$ and $\Delta B$ term, 
\begin{equation}
-\frac{1}{2} \int dr dr^{\prime}\left( -\frac{B\Delta}{\pi^2} \left\langle \cos(\sqrt{4\pi}\phi(r)) \cos(\sqrt{4\pi}\theta(r^{\prime}))\right\rangle 
- \left\langle \cos(\sqrt{4\pi}\phi(r)) \right\rangle \left\langle \cos(\sqrt{4\pi}\theta(r^{\prime}))\right\rangle \right) = 0.
\end{equation}
Now we rescale the first and second order cumulant expansions by replacing $r=br^{\prime}$, $\omega=\frac{\omega^{\prime}}{b}$, $\phi_s(r)=\phi^{\prime}(r^{\prime})$ and $\phi(\omega)=b\phi^{\prime}(\omega^{\prime})$.
\begin{equation}
\begin{split}
\left\langle S_{int}(\phi,\theta) \right\rangle_f=b^{2-K}\int dr^{\prime} \left( \frac{B}{\pi} \cos(\sqrt{4\pi}\phi^{\prime}(r^{\prime}))\right) - b^{2-\frac{1}{K}}\int dr^{\prime} \left( \frac{\Delta}{\pi}\cos(\sqrt{4\pi}\theta^{\prime}(r^{\prime}))\right), 
\end{split}
\end{equation}
\begin{dmath}
-\frac{1}{2}(\left\langle S_{int}^2\right\rangle-\left\langle S_{int}\right\rangle^2) 
= \frac{B^2}{4\pi^2}\left(b^2- b^{2-2K}\right)\int dr^{\prime} (\partial_{r^{\prime}}\phi^{\prime})^2 - \frac{B^2}{2\pi^2}\left(b^{2-4K}- b^{2-2K}\right)\int dr^{\prime} \cos[\sqrt{16\pi}\phi^{\prime}(r^{\prime})] + \frac{\Delta^2}{4\pi^2}\left(b^2- b^{2-\frac{2}{K}}\right)\int dr^{\prime} (\partial_{r^{\prime}}\theta^{\prime})^2 
-\frac{ \Delta \mu}{\pi^2 v}(b^2-b^{2-\frac{1}{K}})\int dr^{\prime} \cos[\sqrt{4\pi}\theta^{\prime}(r^{\prime})]. 
\end{dmath}
Comparision of $B$ terms,
\begin{equation*}
B^{\prime}=Bb^{2-K}. 
\end{equation*}
We put $b=e^{dl}$ and expand the exponential upto second term, i.e, $e^{dl}=1+dl$. Then,
\begin{equation*}
\begin{aligned}
B^{\prime}&=B[1+(2-K)dl] 
&=B+(2-K)Bdl. 
\end{aligned}
\end{equation*}
We define $B^{\prime}-B=dB$, thus we have,
\begin{equation}
\frac{dB}{dl}=(2-K)B. 
\end{equation}
Similarly on comaprision of $\Delta$ terms we have,
\begin{equation}
\frac{d\Delta}{dl}=\left[2-\frac{1}{K}\left( 1+\frac{\mu}{v\pi}\right)  \right] .
\end{equation}
Comparision for $K$ terms gives,\\
\begin{equation}
\frac{dK}{dl}=\frac{1}{2\pi^2}(\Delta^2-B^2K^2). 
\end{equation}
Thus we obtain RG equations,
\begin{equation}
\frac{dB}{dl}=(2-K)B, 
\end{equation}
\begin{equation}
\frac{d\Delta}{dl}=\left[2-\frac{1}{K}\left( 1+\frac{\mu}{v\pi}\right)
\right],
\end{equation}
\begin{equation}
\frac{dK}{dl}=\frac{1}{2\pi^2}(\Delta^2-B^2K^2). 
\end{equation}
Now we consider the limits $B=0$ and $\Delta=0$ to obtain the two BKT equations. Considering $B=0$ we get,
\begin{equation}
\frac{d\Delta}{dl}=\left[2-\frac{1}{K}\left( 1+\frac{\mu}{v\pi}\right)
\right], \;\;\;\;\;\;\;\;
\frac{dK}{dl}=\Delta^2.
\end{equation}
This is the RG equation for the Hamiltonian $H_3$ in eq.\ref{withmu}. Considering $\Delta=0$ we get,
\begin{equation}
\frac{dB}{dl}=(2-K)B,  \;\;\;\;\;\;\;\;
\frac{dK}{dl}=-B^2K^2.
\end{equation}

\end{document}